%% file: main.tex
\documentclass[a4paper,11pt]{article}
\usepackage{jheppub} 

\usepackage{slashed}
\usepackage{lineno}
\usepackage{graphicx}
\usepackage{lipsum}
\usepackage{slashed}
\usepackage{amsmath}
\usepackage[normalem]{ulem}
\usepackage{enumitem}\usepackage{mathtools}
\usepackage{bm}
\usepackage[all]{hypcap}
\usepackage{bbm}
\usepackage{xcolor}
\usepackage{subcaption}
\usepackage{mathabx}
\usepackage{pifont}
\usepackage{multirow}
\usepackage{tablefootnote}
\newcommand{\cC}{\mathcal{C}}

\newcommand{\cB}{\mathcal{B}}
\title{Flavored Circular Collider: cornering New Physics at FCC-ee via flavor-changing processes}

\author[a]{Lukas Allwicher,} 
\emailAdd{lukas.allwicher@desy.de}

\author[b]{Gino Isidori,} 
\emailAdd{isidori@physik.uzh.ch}

\author[b]{and Marko Pesut}
\emailAdd{marko.pesut@physik.uzh.ch}

\affiliation[a]{Deutsches Elektronen-Synchrotron DESY, Notkestr. 85, 22607 Hamburg, Germany}

\affiliation[b]{Physik-Institut, Universit\"at Z\"urich, 8057 Z\"urich, Switzerland}

\abstract{We illustrate the potential of a future high-intensity $e^+ e^-$
 collider running at the $Z$ pole in probing extensions of the Standard Model via  precise measurements of flavor-changing processes. We illustrate this potential both within effective field theories and simplified models inspired by current $B$-physics data, focusing on selected flavor-physics measurement projections at  FCC-ee, and by the theoretically well-motivated scenario of TeV-scale new physics predominantly coupled to third-generation fields. In particular, we demonstrate the key role played by the interplay among different flavor-physics measurements, and between flavor and electroweak measurements, in cornering the New Physics parameter space. Updated constraints on new physics, in the limit that no deviations from the Standard Model are observed, are also presented.}

\begin{document}
\preprint{ZU-TH 17/25, DESY-25-046}

\maketitle 

\section{Introduction}

Flavor physics represents one of the most intriguing aspects of the Standard Model (SM) and, at the same time, provides a powerful tool for exploring physics beyond the SM. On the one hand, the hierarchical structure of quark and charged-lepton mass matrices 
may be the indication of new flavor non-universal dynamics at high energies. On the other hand, the strong suppression of several flavor-changing processes within the SM enables us to probe, via precision studies of such processes, dynamics occurring at energies that extend significantly above the electroweak scale.  

The potential for discoveries in this sector remains very high. Recent studies have shown that models in which the breaking of the flavor degeneracy is not only due to the SM Yukawa couplings, but which feature a structural difference between third-generation fermions and light fermions, are particularly interesting~\cite{Davighi:2023iks, Allwicher:2023shc, Glioti:2024hye}. Within these models, further assuming that the lowest-energy layer of New Physics (NP) interacts predominantly with third-generation fermions allows for both minimizing the tuning associated with the electroweak hierarchy problem and providing a natural justification for the observed fermion hierarchies. 
Unsurprisingly, some of the strongest constraints on this well-motivated class of models are derived from $b$ and $\tau$ physics, and detectable deviations from SM predictions in these processes could be within the reach of future experiments. 

An $e^+e^-$ collider running at the $Z$ pole with high intensity, such as the FCC-ee proposal at CERN~\cite{FCC:2018evy}
or the CEPC project in China~\cite{CEPCStudyGroup:2018ghi}, could offer unprecedented improvements of several key flavor-physics measurements,
especially concerning $b$'s and $\tau$'s. 
Although specific studies of flavor-changing processes at these colliders already exist (see, e.g.,~\cite{Monteil:2021ith,Amhis:2023mpj,Zuo:2023dzn,Ai:2024nmn} and references therein),
it is fair to state that the overall potential of the corresponding measurements is still largely unexplored. A particularly noteworthy aspect is the interplay among different flavor-physics measurements, and between flavor and electroweak measurements, within the same experimental setup. The purpose of this paper is to highlight this potential via the {\em combined} analysis of selected flavor-physics measurements also in conjunction with electroweak observables. 

To illustrate this point, we focus on explicit examples where flavor-physics observables, and in particular the complementarity of different measurements, play a key role in constraining the underlying model. For concreteness, and also 
for their theoretical interest, we focus on models where NP is  coupled mainly to third generation fermions and is characterized by  an approximate $U(2)^5$ flavor symmetry~\cite{Barbieri:2011ci,Isidori:2012ts,Faroughy:2020ina}. As a follow-up of the general Effective Field Theory (EFT) analysis  of such models presented in Ref.~\cite{Allwicher:2023shc}, we analyze the improvements on NP bounds expected from flavor observables in the limit where no deviations from the SM are detected. We do this by taking into account the potential impact of FCC-ee, which we adopt as a reference for the future $Z$-boson factory, compared with the improvements expected from other experiments (mainly LHCb and Belle-II) over the next 20 years. In order to illustrate the complementarity of different measurements, we also analyze in detail some benchmark points where clear deviations from the SM could be detected. To define such benchmark points, we start from the current mild tensions between data and observations in semileptonic B decays (see e.g.~\cite{Allwicher:2024ncl}). We analyze these hypothetical signals not only within the EFT approach but also within explicit simplified  models, thereby illustrating interesting differences in determining the underlying theory in the two cases.

The paper is organized as follows: in Section~\ref{sect:strategy}
we introduce the general ingredients of the whole analysis,  
namely the observables we consider, their expected improvements both before and after FCC-ee,
and the general treatment of NP effects. In Section~\ref{sect:EFT} we analyze the NP benchmarks related to NP coupled mainly to the third generation within the EFT approach and in Section~\ref{sect:UV} using explicit simplified models. In Section~\ref{sect:limits} we present the improved limits on $U(2)^5$ invariant EFT operators from future flavor measurements in the absence of deviations from the SM.
The results are summarized in the Conclusions. 
Explicit expressions for NP contributions to the observables we consider are reported in Appendix~\ref{app:Flavobs}.

\begin{table}[t]
    \centering
    \resizebox{\textwidth}{!}{
    \begin{tabular}{|c|c|c|c|c|}
        \hline
        Observable & SM & Current value~\cite{ParticleDataGroup:2024cfk}  & Pre-FCC projection & FCC-ee expected \\ \hline\hline
        $\left|g_\tau/g_\mu\right|$ & 1 & $1.0009 \pm 0.0014$ & -- & 
        $\pm 0.0001$~\cite{Lusiani:2023tau} \\
        $\left|g_\tau/g_e\right|$ & 1 & $1.0027 \pm 0.0014$ & --  & 
        $\pm 0.0001$~\cite{Lusiani:2023tau}\\
        corr. & & 0.51 & & \\ \hline
        $\cB(\tau \to \mu\bar\mu\mu)$ & 0 &
        $ < 2.1 \times 10^{-8} $         
        & 
        $  < 0.37 \times 10^{-8}$~[*]~\cite{Belle-II:2018jsg} 
        & 
        $ < 1.5 \times 10^{-11}$~[*]~\cite{Lusiani:2023tau}
        \\ \hline
        $R_D$ & $0.298 \pm 0.004$ & $0.342 \pm 0.026$~\cite{HeavyFlavorAveragingGroupHFLAV:2024ctg} & $\pm 3.0\%$~\cite{Belle-II:2018jsg}  & \\
        $R_{D^*}$ & $0.254 \pm 0.005$ & $0.287 \pm 0.012$~\cite{HeavyFlavorAveragingGroupHFLAV:2024ctg} & $\pm 1.8\%$~\cite{Belle-II:2018jsg} & \\ 
        corr. & & -0.39 &   & \\ 
        $\cB(B_c \to \tau\bar\nu)$ & $(1.95 \pm 0.09)\times 10^{-2}$ & $<0.3$ (68\%C.L.) & -- & $\pm 1.6 \% $~\cite{Zuo:2023dzn}\\ \hline
        $\cB(B\to K \nu\bar\nu)$ & $(4.44 \pm 0.30)\times 10^{-6}$ & $(1.3 \pm 0.4)\times 10^{-5}$ & $\pm 14\%$~\cite{Belle-II:2018jsg}  & $\pm 3\%$~\cite{Amhis:2023mpj}  \\
        $\cB(B\to K^* \nu\bar\nu)$ & $(9.8 \pm 1.4)\times 10^{-6}$ & $<1.2 \times 10^{-5}$ (68\%C.L.) & $\pm 33\%$~\cite{Belle-II:2018jsg} & $\pm 3\%$~\cite{Amhis:2023mpj}   \\ \hline
        $\cB(B\to K \tau\bar\tau)$ & $(1.42 \pm 0.14)\times 10^{-7}$ & $<1.5 \times 10^{-3}$ (68\%C.L.) & $<2.7\times 10^{-4}$ &  $\pm 20\%$~[**]~\cite{Miralles:2024iii}  \\
        $\cB(B\to K^* \tau\bar\tau)$ & $(1.64 \pm 0.06)\times 10^{-7}$ & $<2.1 \times 10^{-3}$ (68\%C.L.)  & $< 6.5 \times 10^{-4}$~[*]~\cite{Belle-II:2018jsg}   &  $\pm 20\%$~[**]~\cite{Miralles:2024iii}   \\
        $\cB(B_s\to\tau\bar\tau)$ &  $(7.45 \pm 0.26)\times 10^{-7}$ & $<3.4 \times 10^{-3}$ (68\%C.L.) & $< 4.0 \times 10^{-4}$~[*]~\cite{Belle-II:2018jsg}  &  $\pm 10\%$~[**]~\cite{Miralles:2024iii} \\ \hline
        $\Delta M_{B_s}/\Delta M^{\rm SM}_{B_s}$ &  1 & 
        $\pm 7.6\%$  &  $\pm 3.3\%$~\cite{Charles:2020dfl} &  $\pm 1.5\%$~\cite{Charles:2020dfl} \\ \hline
        $\cB(B\to K \tau\bar\mu)$ & 0 & &$< 1.0 \times 10^{-6}$~[*]~\cite{LHCb:2018roe}  & \\
        $\cB(B_s\to\tau\bar \mu)$ & 0 & &$< 1.0 \times 10^{-6}$~[*]~\cite{LHCb:2018roe}  & \\ \hline
        \hline
    \end{tabular}}
    \caption{ \label{tab:exp} List of the flavor observables we consider in our analysis, with corresponding SM predictions, current 
    experimental values, and expected future sensitivities before the start of FCC-ee and after its completion (see text for more details). 
    The entries marked with [*] are upper bounds in the absence of a signal; the entries marked with [**] are relative errors assuming an enhanced rate over the SM 
    expectation (by a factor $\gtrsim 3$); the other entries are relative errors assuming the SM value.  }
\end{table}

\section{Observables and analysis strategy}
\label{sect:strategy}

As anticipated, the $Z$-pole run of a future circular collider would enable a tremendous advancement in precision flavor physics, especially concerning $b$'s and $\tau$'s. 
For the reference 
figure of ${6} \times 10^{12}$ $Z$ boson at FCC-ee~\cite{FCC:2018evy}, 
one expects $9\times10^{11}$~$b\bar b$  and
$2\times10^{11}$~$\tau^+\tau^-$ pairs produced back-to-back in a relatively clean environment.
The  environment resembles that of the $B$ factories, with three main advantages: 1) a large statistical gain; 2) the possibility to produce all type of $b$-hadrons; 3)~a large boost, which leads to higher efficiency compared to $B$ factories for modes with missing energy, such as di-neutrino or di-tau modes.

In order to illustrate this potential we have selected a series of complementary observables reported in Table~\ref{tab:exp}. This list is far from being exhaustive, but it is sufficient for our illustrative purposes. In the third column of the table we list the current central values of the observables reported by PDG~\cite{ParticleDataGroup:2024cfk} or 
HFLAV~\cite{HeavyFlavorAveragingGroupHFLAV:2024ctg}. 
In the fourth column we indicate the 
expected improvement in precision over the next 20 years, before the start of FCC-ee,  according to the 
upgrade-II of LHCb~\cite{LHCb:2018roe}  and the 
projections of Belle-II~\cite{Belle-II:2018jsg}
assuming an integrated luminosity of 10~ab$^{-1}$ (current plan in absence of a major upgrade). 
Finally, in the last column we report the expected projections at FCC-ee: those are based either on the results of the explicit studies in~\cite{Monteil:2021ith,Amhis:2023mpj,Zuo:2023dzn,Ai:2024nmn} or, for the channels where these are not available, by na\"ive extrapolations of these 
results.\footnote{In the di-neutrino modes we conservatively increase 
the  error for the FCC-ee projections estimated in~\cite{Amhis:2023mpj} 
to 3\%, in order to take into account theoretical uncertainties in the SM predictions.
The projections for the di-tau modes are based on  the results in~\cite{Miralles:2024iii}, extrapolated to all decay modes and assuming enhanced rates, by at least a factor of 3, which is what occurs in all the explicit NP models we are considering.}
The projections for $\tau$ physics are based on Ref.~\cite{Lusiani:2023tau}, while those for $B_s$~mixing are obtained from~\cite{Charles:2020dfl}.

Since one of our main goals is to analyze the complementarity of flavor and electroweak observables, we also consider FCC-ee projections for the latter. 
These are based on the figures quoted in Ref.~\cite{deBlas:2022ofj}, updating the statistical errors to account for 1)~four Interaction Points (IPs) instead of two, and 2)~the latest estimate for the expected integrated luminosity of the $Z$-pole run, namely 205 ab$^{-1}$.
For definiteness, we summarise the electroweak observables we consider and their projected relative uncertainty in Table \ref{tab:ewprojections}.
\begin{table}[]
    \centering
        \begin{tabular}{|c|c||c|c|}
         \hline
         Observable & Relative uncertainty & Observable & Relative uncertainty \\
         \hline\hline 
         $\Gamma_Z$ & $1.0 \times 10^{-5}$  
                & $A_b$ & $2.3 \times 10^{-4}$ \\ 
         $\sigma_{\text{had}}^{0}$ & $9.6 \times 10^{-5}$    
                & $A_\tau$ & $1.4 \times 10^{-3}$ \\ \cline{3-4}
         $R_b$ & $3.0 \times 10^{-4}$  
                & $m_W$ & $4.6\times 10^{-6}$ \\
         $R_\mu$ & $5.0 \times 10^{-5}$        
                & $\Gamma_W$ & $5.1 \times 10^{-4}$  \\
         $R_e$ & $3.0 \times 10^{-4}$  
                & $\cB(W\to \tau\nu)$ & $3.0\times10^{-4}$ \\\cline{3-4}
         $R_\tau$ & $1.0 \times 10^{-4}$ 
                & $\mu(H \to b\bar b)$ & $3.0\times 10^{-3}$ \\ 
         $N_{\rm eff}$ & $0.6 \times 10^{-3}$ 
                &  $\mu(H \to \tau\bar \tau)$ & $9.0\times 10^{-3}$ \\
         \hline
    \end{tabular}
    \caption{Expected relative uncertainties for the relevant EWPOs at FCC-ee used in our analysis. For $Z$- and $W$-decay observables, the numbers are taken from \cite{deBlas:2022ofj}, rescaled for 4 IPs and 205 ab$^{-1}$ integrated luminosity ($Z$ pole). The projection for the effective number of neutrinos $N_{\rm eff}$ is taken from \cite{Blondel:2021ema} and adapted similarly. The projection for Higgs signal strengths follows \cite{Bernardi:2022hny}.
    \label{tab:ewprojections} }
\end{table}

\subsection{Main New Physics hypotheses}

We work under the general assumption of an ultraviolet (UV) completion of the SM 
characterized by new degrees of freedom  above the electroweak scale. Under this assumption, deviations from the SM in $b$ and $\tau$ decays, as well as in electroweak observables,  are accurately described  by matching the explicit NP model into the SM Effective Field Theory (SMEFT)~\cite{Buchmuller:1985jz,Grzadkowski:2010es,Brivio:2017vri,Isidori:2023pyp},
considering effective operators up to dimension six, and then evaluating the  observables within the SMEFT.
Employing the so-called Warsaw basis~\cite{Grzadkowski:2010es} and using the electroweak scale as normalization for the dimensionless Wilson coefficients, the SMEFT 
Lagrangian assumes the form
\begin{align}
    \mathcal{L}_{\rm SMEFT} = \mathcal{L}_{\rm SM} - \frac{2}{v^2} \sum_k \cC^{[f]}_k Q_k \,
    \label{eq:Leff}
\end{align}
where $v = (\sqrt{2}G_F)^{-1/2}\approx 246$~GeV.
Here $k$ denotes the electroweak structure and field content of the operators, and $[f]$ their flavor structure, following the convention of Ref.~\cite{Allwicher:2023shc}.
By default, for the quark doublets ($q^i$) we adopt the down-quark 
mass-eigenstate basis ($i\in\{1,2,3\}\equiv\{d,s,b\}$), while lepton doublets are understood to be in the charged-lepton mass-eigenstate basis. 
The explicit expressions for the observables we consider in terms of the $\cC^{[f]}_k$ 
are reported in Appendix~\ref{app:Flavobs}.

Besides the general hypothesis of heavy NP, we also assume that NP is coupled mainly to third-generation fermions and is characterized by an approximate $U(2)^5$ flavor symmetry acting on the light generations~\cite{Barbieri:2011ci,Isidori:2012ts,Faroughy:2020ina}. This more specific hypothesis finds a natural root in models where the hierarchical pattern of SM Yukawa couplings is the  result of a multi-scale structure~\cite{Dvali:2000ha,Panico:2016ull,Barbieri:2021wrc,Allwicher:2020esa}, possibly associated to a flavor non-universal gauge group~\cite{Allwicher:2020esa,Bordone:2017bld,Greljo:2018tuh,Fuentes-Martin:2020pww,Fuentes-Martin:2022xnb,Davighi:2022bqf,
Davighi:2023iks, Davighi:2023evx,Barbieri:2023qpf,
FernandezNavarro:2023rhv,FernandezNavarro:2024hnv}. 
A complementary motivation to explore this class of UV completions comes from the experimental bounds on NP: they are particularly stringent for processes involving light families, both from direct searches and from flavor observables, while they do not exceed 1-2 TeV for particles coupled only to the third generation.  TeV-scale NP dominantly coupled to third-generation fields is therefore the best option to minimize the little hierarchy problem associated to the Higgs sector, as long as its couplings to the first and second families are sufficiently small and flavor universal~(see Ref.~\cite{Fuentes-Martin:2020pww,Fuentes-Martin:2022xnb,Stefanek:2024kds,Covone:2024elw} for explicit examples).

In Section~\ref{sect:limits} we analyze the bounds on this class of models in a general bottom-up  approach, under the assumptions that no deviations from the SM will be detected at FCC-ee. 
More precisely, we derive bounds on the Wilson coefficients of $U(2)^5$-invariant  effective operators, both in a down-aligned and in a up-aligned basis for the definition of the $U(2)^5$ symmetry. As stated in the introduction, we are also interested in analyzing explicit NP frameworks that illustrate the interplay between different observables, as well as discussing the impact of the leading breaking of the $U(2)^5$ symmetry, which plays a key role in flavor observables. To this purpose, we need to restrict the focus on specific parameter regions of the EFT (Section~\ref{sect:EFT}) and specific UV completions (Section~\ref{sect:UV}). We do so taking into account the tensions 
presently observed in $B$-physics data.

\subsection{$U(2)^5$ breaking and quark flavor mixing}

Despite being a good approximate symmetry, $U(2)^5$ is necessarily broken, at least in the Yukawa sector, and its breaking has a crucial impact in flavor physics. We work under the assumption of a minimal $U(2)^5$ breaking,  as originally defined in~\cite{Barbieri:2011ci} in the quark sector. 
In this limit, the dominant breaking is the spurion  $\tilde{V}_q$, 
transforming as a doublet of  $U(2)_q$, which is responsible for the heavy$\to$light mixing in the Cabibbo-Kobayashi-Maskawa (CKM)
matrix. The  spurion is parameterized as
\begin{equation}
  \tilde{V}_q=  \begin{pmatrix}
     V_{td}  \\[1.5mm]
     V_{ts} 
    \end{pmatrix}\,,
    \label{eq:Vform}
\end{equation}
where $V_{td}$ and $V_{ts}$ are the elements of the CKM matrix.
 We neglect subleading spurions related to light quark masses 
(i.e.~we work in the limit $m_s=m_d=0$).

In the $U(2)^5$ setup there is an intrinsic ambiguity on what we denote as third generation in the left-handed quark sector, or alternatively, which are the $U(2)_q$ singlet fields. Starting from the  down-aligned  basis that we use as reference, where the $U(2)_q$ singlet field is $q^3 \equiv q^b$,
we can  span equivalent bases via appropriate insertions of the spurion $\tilde{V}_q$:
\begin{eqnarray}
q^{3\prime} &=&  q^3  - \varepsilon  \tilde{V}_q^i  q^i~.
\label{eq:epsf}
\end{eqnarray}
Varying the $O(1)$ parameter $\varepsilon$ between 0 and -1 we move from the down-aligned basis to the up-aligned one for the $U(2)_q$ singlet field:
\begin{equation}
q^{3\prime}|_{\varepsilon=0}   =\begin{pmatrix}  
    V^*_{ub} u_L + V^*_{cb} c_L + V^*_{tb} t_L \\
    b_L
\end{pmatrix}\,, \quad 
q^{3\prime}|_{\varepsilon=-1} \approx
 \begin{pmatrix}  
     t_L \\
    V_{td} d_L + V_{ts} s_L + V_{tb} b_L
\end{pmatrix}\,, 
\end{equation}
In order to avoid the appearance of new CP-violating phases beyond the CKM one, we assume $\varepsilon$ to be real.
Since in the standard CKM convention
$V_{ts}\approx -0.04$, the sign of $\varepsilon$ in (\ref{eq:epsf}) is such that $\varepsilon>0$ implies a positive mixing of $q^2$ and $q^3$ in $q^{3\prime}$. 

\section{Discovery potential: EFT studies}
\label{sect:EFT}

\subsection{Effective operators}

The starting point of our EFT analysis are the persistent discrepancies between observations and SM predictions in three sets of semileptonic $B$ decays. First, the Lepton Flavor Universality (LFU) ratios $R_D$ and $R_{D^*}$, which indicate a $\sim 3 \sigma$ tension with the SM (see~\cite{Iguro:2024hyk,HeavyFlavorAveragingGroupHFLAV:2024ctg} for a recent discussion). Second, 
 the enhancement with respect to the SM of $B^+\to K^+\nu\bar\nu$~\cite{Belle-II:2023esi} and $K^+ \to \pi^+ \nu\bar\nu$~\cite{NA62:2024pjp} rates. Third, the
determination of the effective vector-current coupling ($C_9$) from $b\to s\ell\bar \ell$ transitions ($\ell=e,\mu$), which is more uncertain but nevertheless indicates a tension with the SM of at least $2\sigma$ (see~\cite{Alguero:2022wkd,Bordone:2024hui,Isidori:2024lng}). None of these effects are particularly convincing on their own. However, when considered together, they form a coherent and interesting picture: all point to non-standard coefficients for semileptonic operators involving only third-generation leptons and quark fields that are also closely aligned with the third generation~\cite{Allwicher:2024ncl}.

These considerations lead us to consider as starting point the following set of 
$U(2)^5$-invariant 
effective operators involving only heavy fields:
\begin{align}
Q_{\ell q} ^{(1)[3333]} & = (\bar{\ell}^3\gamma_{\mu}\ell^3)
(\bar{q}^3 \gamma^{\mu} q^3)\,,\\
Q_{\ell q}^{(3)[3333]} & =(\bar{\ell}^3\gamma_{\mu}\sigma^a\ell^3)
( \bar{q}^3 \gamma^{\mu}\sigma^a q^3)\,,\\
Q^{[3333]}_{\ell e q d} &= (\bar{\ell}^3 e^3) 
( \bar{d}^3  q^3)\,.
\label{eq:ops} 
 \end{align}
To describe heavy$\to$light quark mixing, in addition to these operators, we should consider terms generated by the insertion of at least one spurion, via the replacement 
\begin{equation}
q^3 \to q^{3\prime}  = q^3  - \varepsilon  \tilde{V}_q^i  q^i~.
\label{eq:eps_rep}
\end{equation}
To avoid the proliferation of free parameters, we  assume that
the underlying NP leads to a rank-one structure in quark flavor space~\cite{Marzocca:2024hua}. 
In other words, we assume that NP is aligned to a specific direction in flavor space identified by the definition of the $U(2)_q$ singlet field $q^{3\prime}$ in (\ref{eq:epsf}). 
This direction is neither the down-aligned nor the up-aligned basis (defined by the diagonalization of the corresponding Yukawa couplings), but is a new direction characterized by a specific value of $\varepsilon$. Under this assumption, we can describe NP effects via the three operators in (\ref{eq:ops}) replacing all the $q^3$ fields with $q^{3\prime}$, thereby obtaining a universal relation between operators with $q^3$ and $q^{1,2}$ fields controlled only by  the value of~$\varepsilon$.

\subsection{Definition of the NP benchmarks}
\label{Sect:EFT_bench}

\begin{figure}[t]
\centering
\includegraphics[width = 7.3cm]{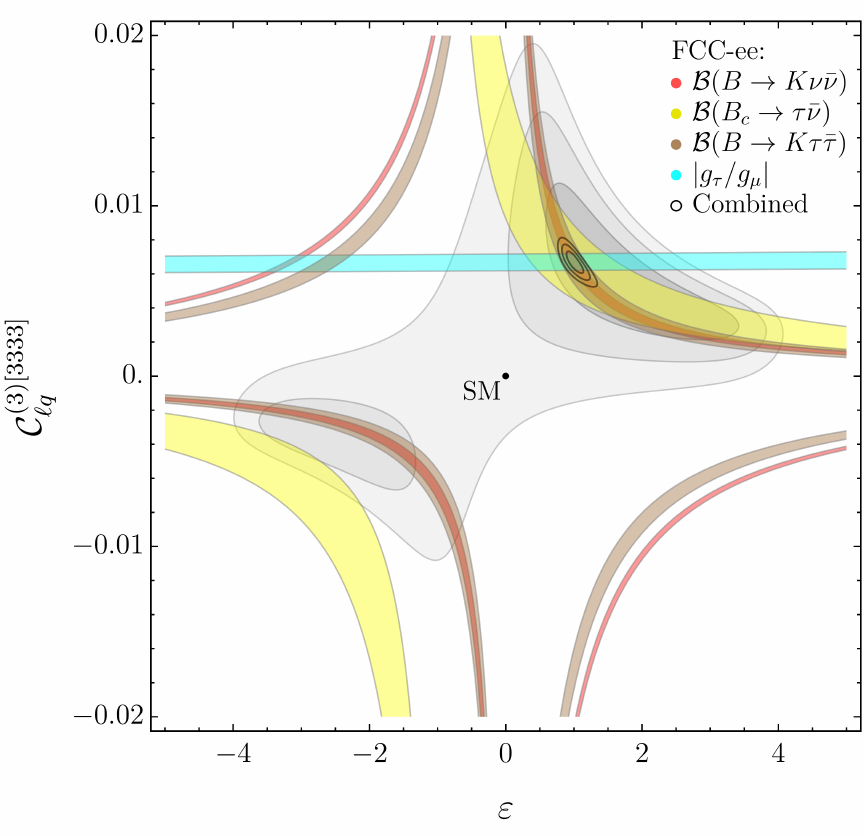}
\hskip 0.3 cm
\includegraphics[width = 7.3cm]{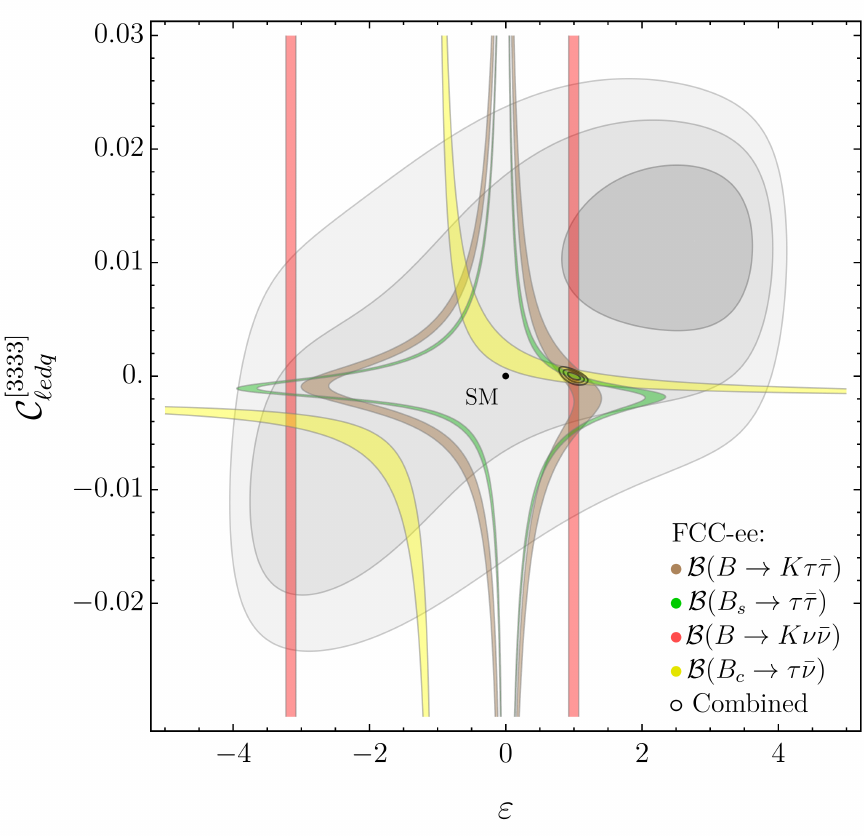}
\caption{Present and future constraints in the $\{ \varepsilon, \cC_{\ell q}^{(3)[3333]} \}$
 and  $\{ \varepsilon, \cC_{\ell e d q}^{[3333]} \}$ planes.
The gray areas are the regions allowed at 1, 2, and $3\sigma$ by present constraints. 
The colored bands denote the possible future impact of specific flavor observables at FCC-ee, 
at $1\sigma$, assuming the EFT NP benchmark I. The result of the global fit at FCC-ee from flavor and electroweak observables are indicated by the small ellipses (corresponding to 
1, 2, and $3\sigma$ intervals).
\label{Fig:EFT1}}
\vskip 0.5 cm
\includegraphics[width = 7.3cm]{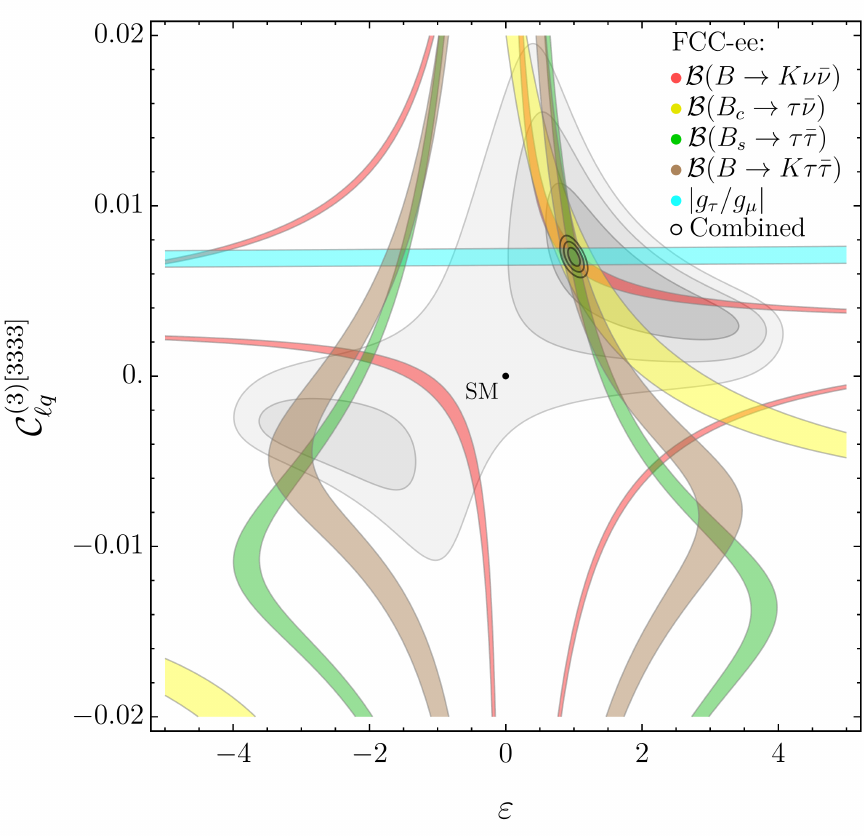}
\hskip 0.3 cm
\includegraphics[width = 7.3cm]{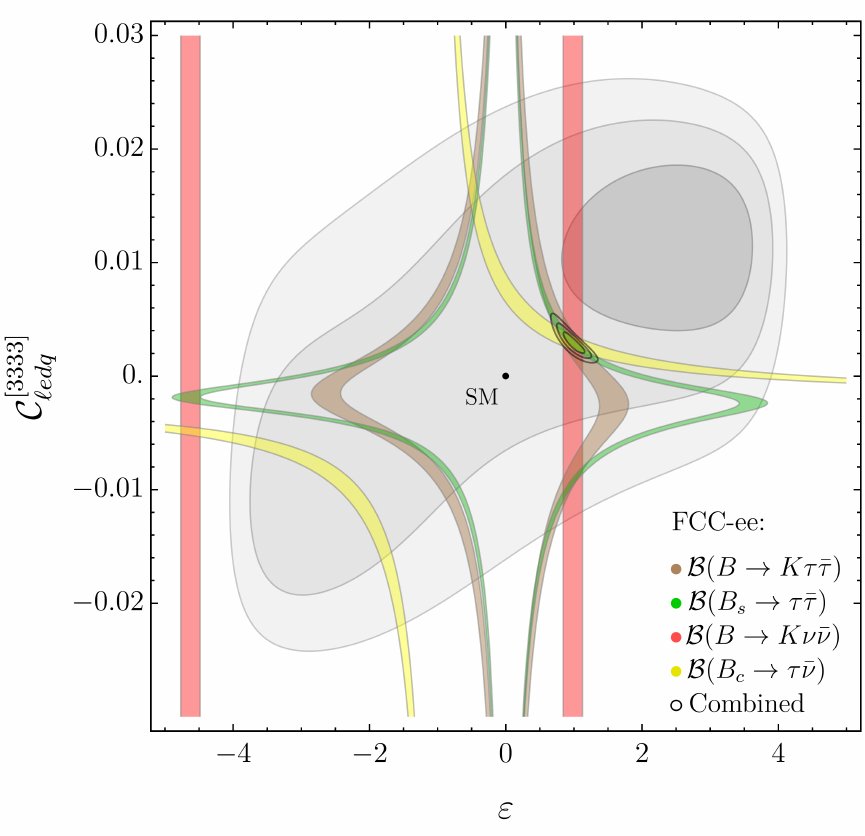}
\caption{Present and future constraints in the $\{ \varepsilon, \cC_{\ell q}^{(3)[3333]} \}$
  and $\{ \varepsilon, \cC_{\ell e d q}^{[3333]} \}$ planes, 
assuming the EFT NP benchmark II. Notations as in Fig.~\ref{Fig:EFT1}.
\label{Fig:EFT2}}
\end{figure}

We are considering an EFT characterized by four parameters, the coefficients of the three operators in (\ref{eq:ops}) and the value of~$\varepsilon$. At present, 
a global (four-dimensional) fit of these parameters give rise to different, 
almost degenerate, minima. A better illustration of the presently allowed range
is obtained setting some of the Wilson coefficients to zero. The results thus 
obtained in the $\{ \varepsilon, \cC_{\ell q}^{(3)[3333]} \}$
and $\{ \varepsilon, \cC_{\ell e q d}^{[3333]} \}$ planes are shown in Fig.~\ref{Fig:EFT1} (or, equivalently, in Fig.~\ref{Fig:EFT2}). Here the gray areas are determined both by the current flavor constraints\footnote{For a more conservative estimate of the allowed range we do not include data from $b\to s\ell\bar \ell$ transitions, which are subject to sizable theoretical uncertainties.} 
in Table~\ref{tab:exp} and by the constraints from electroweak and collider data (for which we refer to Ref.~\cite{Allwicher:2023shc}). As can be seen, present data hint to non-zero positive values of $\cC_{\ell q}^{(3)[3333]}$ and  $\varepsilon$, but are compatible with the SM within 
$3\sigma$. 

To illustrate the  impact of flavor measurements at FCC-ee we consider the following two NP benchmarks:
\begin{itemize}
\item[I.] $\cC_{\ell q}^{(3)[3333]} =6.7 \times 10^{-3}$, $\varepsilon=1$,  
$\cC_{\ell q}^{(1)[3333]}= \cC_{\ell e q d}^{[3333]} =0,$
\item[II.] $\cC_{\ell q}^{(3)[3333]} =6.7 \times 10^{-3}$, $\varepsilon=1$,  
$\cC_{\ell q}^{(1)[3333]}= \cC_{\ell e q d}^{[3333]} =3.0 \times 10^{-3}$,
\end{itemize}
which both belong to the $1\sigma$ allowed region of the current global (four-dimensional) fit.

\begin{figure}[t]
\centering
\includegraphics[width = 8.5 cm]{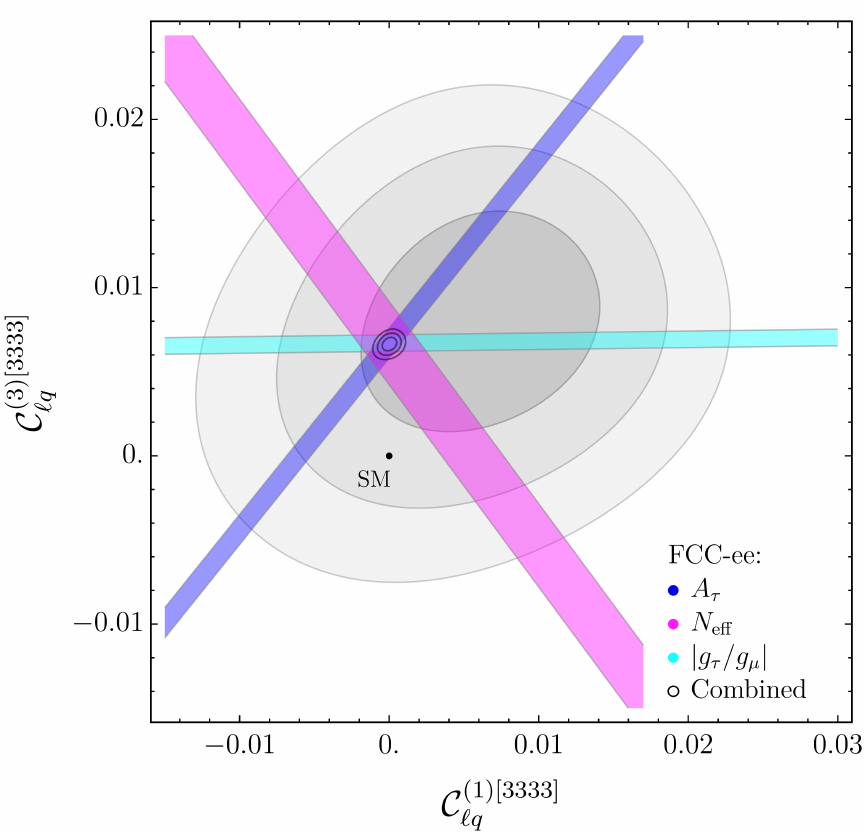}
\caption{Present and future constraints in the $\{  \cC_{\ell q}^{(1)[3333]} , \cC_{\ell q}^{(3)[3333]} \}$ plane, assuming the EFT NP benchmark I. Notations as in Fig.~\ref{Fig:EFT1}.
\label{Fig:EFT3}}
\end{figure}

In  Fig.~\ref{Fig:EFT1}, \ref{Fig:EFT2}, and \ref{Fig:EFT3} we illustrate how future measurements at FCC-ee, from both flavor and electroweak observables, would allow us to unambiguously identify and distinguish these different NP scenarios. 
All the projected FCC-ee measurements are obtained assuming, as central value, the expectation for the observable in the respective NP benchmark and, as relative error the one reported in the last column of Table~\ref{tab:exp}.
All plots are the result of four-dimensional fits of the NP parameters, where the two parameters not shown in the plane are set to their best-fit value. 

Fig.~\ref{Fig:EFT1} and \ref{Fig:EFT2} elucidate the role of flavor observables, which are clearly  essential to determine the value of $\varepsilon$. It is interesting to note how the constraints from a given flavor observable change from Fig.~\ref{Fig:EFT1} to \ref{Fig:EFT2}, reflecting the dependence on the parameters not shown in the plot. In the case of $|g_\tau / g_\mu|$, which depends only on $\cC_{\ell q}^{(3)}$ and  $\varepsilon$, there is no change, while observables such as 
$\cB(B_s \to \tau^+\tau^-)$ and $\cB(B \to K \tau^+\tau^-)$, which are highly sensitive to scalar amplitudes (and thus to $\cC_{\ell e d q}$), exhibit a large variation. This comparison nicely illustrates one of the major virtues of FCC-ee: it allows the determination of several independent high-precision observables. In indirect NP searches, the combination of different observables is essential both to break degeneracies in the NP parameter space and to corroborate the evidence of non-SM effects. 

The complementarity of flavor and electroweak observables is illustrated by 
Fig.~\ref{Fig:EFT3}. We produce this figure only for one of the benchmarks since the two cases are very similar. Note in particular the interplay of $N_{\rm eff}$ and $A_\tau$
in resolving the $SU(2)_L$ structure of the EFT. As in the case of 
Fig.~\ref{Fig:EFT1} to \ref{Fig:EFT2},
only a subset of the relevant observables
is explicitly shown. 
 
To conclude this section, it is worth stressing that the NP benchmarks we have analyzed correspond to high-scale physics fully compatible with present direct searches (which are taken into account when determining the current allowed region). More precisely, given the normalization of  the effective Lagrangian in (\ref{eq:Leff}), the value of $\cC_{\ell q}^{(3)[3333]}$ in the two benchmarks corresponds to an effective scale
$\Lambda  = v /\sqrt{2 \cC_{\ell q}^{(3)[3333]} } \approx 2.1$~TeV for the 
corresponding effective operator.

\section{Discovery potential: simplified  models}
\label{sect:UV}

We now turn our attention to simplified UV models which can realize, at least in part,  
the EFT benchmarks for the semileptonic operators analyzed above. The interest of considering concrete models is twofold: first, within explicit models we can connect different sets of observables, such as $\Delta F=2$
or Higgs couplings, which in a pure EFT approach are completely disconnected from semileptonic processes (up to loop- and CKM-suppressed effects). Second, this exercise illustrates some of the challenges in determining the parameter space of a concrete model vs.~a mere determination of the EFT couplings.
For each model, we couple the new fields only to the third-generation SM fields, leaving implicit the fact that flavor-violating effects are obtained by the replacement $q^3 \to q^{3\prime}  = q^3  - \varepsilon  \tilde{V}_q^i  q^i$, with $\varepsilon$ an $\mathcal{O}(1)$ parameter.

\subsection{$U_1$+$Z'$}

The semileptonic operators 
in Eq.~(\ref{eq:ops}) can be efficiently generated, at the tree level, by the exchange of a $SU(2)_L$--singlet vector leptoquark ($U_1$) coupled mainly to the third generation. 
A field with these properties, with a mass in the TeV range, was indeed recognized as a very efficient mediator to explain the $R_{D^{(*)}}$ anomalies while satisfying current experimental constraints~\cite{Alonso:2015sja,Calibbi:2015kma,Barbieri:2015yvd,Bhattacharya:2016mcc,Buttazzo:2017ixm}. Such field does appear as massive gauge boson in models with extended gauge symmetries (see e.g.~\cite{DiLuzio:2017vat,Bordone:2017bld,Greljo:2018tuh}). A key observation is that the $U_1$ boson never appears alone and, being associated to a broken $SU(4)$ group, is always accompanied by one or more massive field coupled to 
a neutral-current, related to the closure of the algebra~\cite{Barbieri:2015yvd,Baker:2019sli}. 

To investigate the correlations between $U_1$--mediated effects and those related to neutral-current mediators, we consider a simplified model consisting of a pair of $U_1$ and $Z'$ massive vector fields, with mass $M_U$ and $M_Z$, respectively, and unsuppressed couplings to third-family fermions. More precisely, we employ the following Lagrangian for the leading interactions,
\begin{align} 
\mathcal{L}_{\text {int }} \supset \,\frac{g_4}{\sqrt{2}} U_\mu \left(\bar{q}^3_L \gamma^\mu \ell^3_{L}\right) +\frac{g_4}{2\sqrt{6}} Z_\mu^{\prime}\left(\bar{q}^3_L \gamma^\mu q^3_L\right)-\frac{3}{2}\frac{g_4}{\sqrt{6}} Z_\mu^{\prime}\left(\bar{\ell}^3_L \gamma^\mu \ell^3_L\right) +\text { h.c. },
\end{align}
where the normalization of the different terms follows from the assumption that the $U_1$ and $Z'$ fields are related to the generators of an $SU(4)$ group \`a la Pati-Salam~\cite{Pati:1974yy}, acting only on the third generation
($Z'$ is associated to the $B-L$ generator). We are aware that this simplified model does not correspond to a fully realistic UV theory, where additional massive bosons do necessarily appear\footnote{Assuming the realistic breaking pattern $SU(4)\times SU(3) \times U(1)_X \to SU(3)_c\times U(1)_{Y}$ leads to an additional electrically-neutral color-octet field~\cite{DiLuzio:2017vat}. 
} and the $Z^\prime$ boson necessarily mixes with the SM $Z$ boson. Still, this simplified model serves the purpose of illustrating the correlations we are interested in.

Integrating out the $U_1$ field leads to the following
tree-level matching conditions for the semileptonic effective operators
\begin{equation}
    \cC_{\ell q}^{(1)[3333]}=  \cC_{\ell q}^{(3)[3333]}=\frac{g_4^2 v^2}{8M_U^2}\,,
    \label{eq:Mod1-CU}
\end{equation}
while integrating out the $Z^\prime$ leads to 
\begin{equation}
  \cC_{\ell q}^{(1)[3333]}=-\frac{g_4^2 v^2}{32M_Z^2}\,, \qquad \cC_{qq}^{(1)[3333]}= \frac{g_4^2 v^2}{192 M_Z^2}\,, \qquad \cC_{\ell\ell}^{[3333]}= \frac{3g_4^2 v^2}{64 M_Z^2}\,.
  \label{eq:Mod1-CZ}
\end{equation}
As expected, the $Z^\prime$ exchange generates not only semileptonic operators but also four-quark and four-lepton  operators. To describe the small, but non-negligible, heavy$\to$light quark mixing, we proceed as 
 in Sect.~\ref{sect:EFT} with the replacement 
 in Eq.~(\ref{eq:eps_rep}) and the hypothesis of a rank-one structure in flavor space. However, we consider different orientations in flavor space for $U_1$ and $Z'$ fields, characterized by different flavor-mixing parameters  
($\varepsilon_{U}$ and $\varepsilon_{Z}$), as expected in realistic models~\cite{DiLuzio:2018zxy}.

\begin{table}[t]
    \centering
    \begin{tabular}{|c|c|c|c|}
        \hline
        Observable & SM   & FCC projection \\ \hline\hline
        $B_c \to \tau\nu$ & $(1.95 \pm 0.09)\times 10^{-2}$ &  $(2.09\pm 0.03)\times 10^{-2}$   \\ 
        $B\to K \nu\bar\nu$ & $(4.44 \pm 0.30)\times 10^{-6}$  & $(5.64 \pm 0.17)\times 10^{-6}$  \\
        $B\to K^* \nu\bar\nu$ & $(9.8 \pm 1.4)\times 10^{-6}$  & $(12.4 \pm 0.4)\times 10^{-6}$   \\ 
        $B\to K \tau\tau$ & $(1.64 \pm 0.06)\times 10^{-7}$  & $(4.2 \pm 0.8)\times 10^{-6}$  \\
        $B_s\to\tau\tau$ &  $(7.45 \pm 0.26)\times 10^{-7}$  & $(2.18 \pm 0.22)\times 10^{-5}$ \\
        $\Delta M_{B_s}/\Delta M^{\rm SM}_{B_s}$ &  1.0 & 
        $0.862\pm 0.015$\\ \hline
         $\left|g_\tau/g_\mu\right|$ & 1.0   & 0.99926(7)   \\ 
        $N_{\rm eff}$ & 3.0   & 2.9979(6)  \\
        $A_\tau$ & $0.147$   & 0.14668(21) \\
        $A_b$ & $0.935$ &   0.93502(22)  \\
        \hline
    \end{tabular}
    \caption{SM values and hypothetical projected values at FCC for the observables we consider in the $U_1$+$Z'$ model.  When analyzing the impact of future measurements, the uncertainties associated to the SM predictions are assumed to be subleading with respect those in the last column.
    \label{tab:U+Z}}
\end{table}

In order to proceed with the analysis, we choose the following  benchmark for the parameters of the model:
\begin{equation}
    \{g_4 = 2\, , \,\, M_U = 2.4\text{~TeV}\, ,  \,\, M_Z = 2\text{~TeV}\, ,  \,\, \varepsilon_U = 2.4\, ,  \,\, \varepsilon_Z = 0.9 \} \,.
    \label{eq:benchM1}
\end{equation}
The values of $g_4$ and $M_U$ are chosen according to two main criteria:~1) they generate coefficients for the semileptonic operators similar to those employed in Sect.~\ref{sect:EFT}, which provide a good description of current $B$-physics data (using the values in (\ref{eq:benchM1}) leads to 
$\cC_{\ell q}^{(3)}\approx0.005$ for third-generation fields);
 2)~they satisfy the constraints from direct searches (see e.g.~\cite{Aebischer:2022oqe}), which forbid too low values for $M_U$.  The value of $M_Z$ is then chosen to be slightly smaller than $M_U$ (in the exact $SU(4)$ limit $M_U=M_Z$), and $\varepsilon_{U,Z}$ are set to be $O(1)$, as expected in realistic models.\footnote{~Note that by means of $\varepsilon_{Z}=O(1)$ we also effectively describe cases where the $Z^\prime$, and other neutral mediators, have no tree-level flavor-violating couplings, but four-quark or four-lepton flavor-violating amplitudes are induced by the leptoquark exchange beyond the tree level~\cite{Fuentes-Martin:2019ign,Fuentes-Martin:2020luw,Fuentes-Martin:2020hvc}.}

The FCC-ee projections  for each observable, $\mathcal{O}$, are obtained assuming, as central value,
\begin{equation}
\left.\mathcal{O}\right|_{\text{FCC-ee}}\equiv \left.\mathcal{O}\right|_{\text{SM}}\left(1+\left.\delta^\text{NP}_{\mathcal{O}}\right|_{\text{Bench. point}}\right)
\end{equation}
where $\left.\delta^\text{NP}_{\mathcal{O}}\right|_{\text{Bench. point}}$ is the NP contribution to the observable evaluated at the benchmark point. The projections thus obtained, with corresponding uncertainties estimated according to Table~\ref{tab:exp}, are  reported in Table~\ref{tab:U+Z}.

\begin{figure}[t]
\centering
\includegraphics[width = 7.3cm]{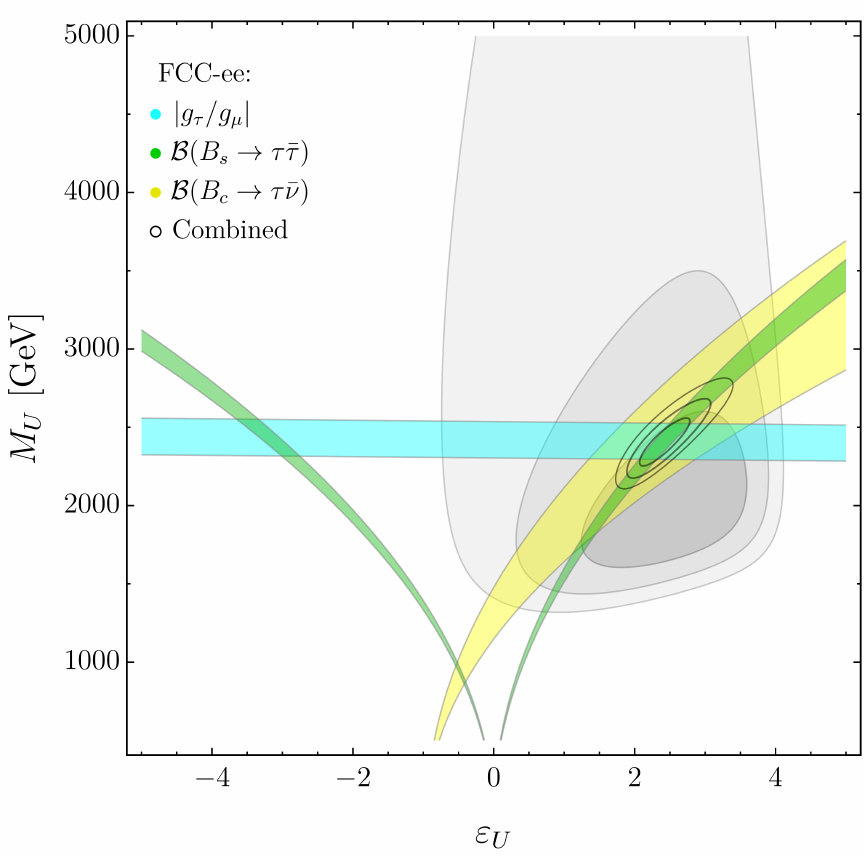}
\hskip 0.3 cm
\includegraphics[width = 7.3cm]{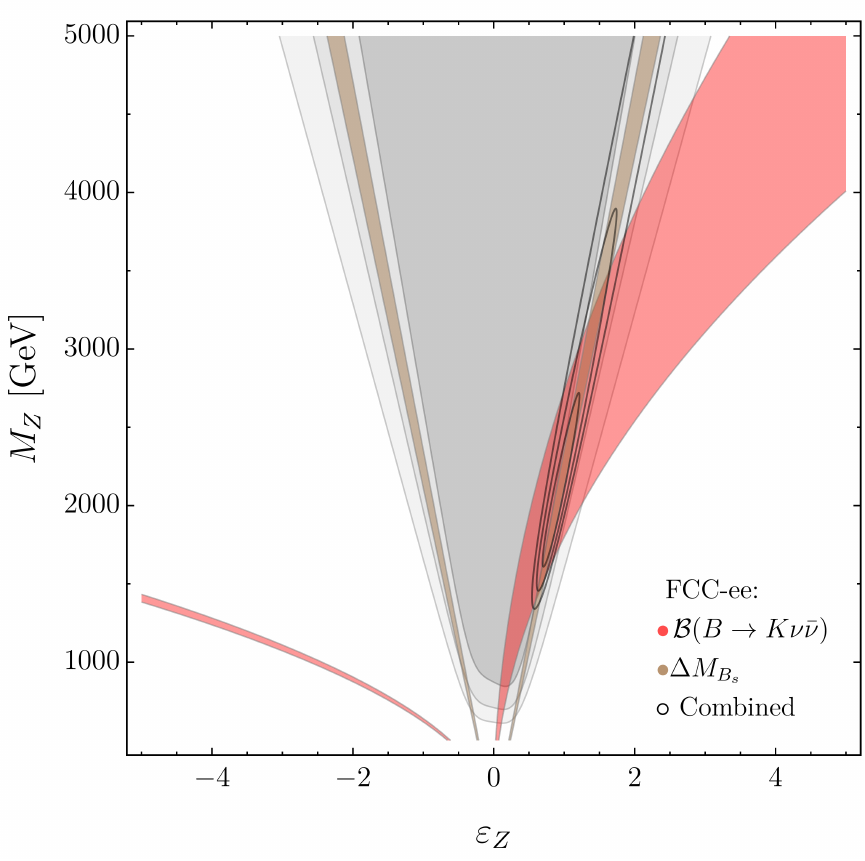}
\caption{Present and future constraints in the $\{ \varepsilon_U, M_U\}$ and 
 and  $\{ \varepsilon_Z, M_Z \}$ planes in the $U_1$+$Z'$ model.
 As in Fig.~\ref{Fig:EFT1},
 the gray areas indicate the regions allowed  by present constraints, the colored bands denote the possible future impact of specific flavor observables at FCC-ee 
 --assuming the NP benchmark in Eq.~(\ref{eq:benchM1})-- and
 the small ellipses indicate the 
 result of the global fit at FCC-ee (in all plots we 
 set $g_4=2$, hence the values of $M_U$ and $M_Z$ need to be rescaled by $g_4/2$ for $g_4\neq 2$).
\label{Fig:model1}}
\end{figure}

In Fig.~\ref{Fig:model1} we illustrate the impact of flavor observables in constraining the model.  Since we analyze only indirect observables, we are sensitive only to the ratios $g_4/M_U$ and $g_4/M_Z$, implying that the actual number of free parameters is four (for convenience we set $g_4=2$,  with the understanding that for $g_4\neq2$ the values of $M_U$ and $M_Z$ need to be properly rescaled).
The left plot in Fig.~\ref{Fig:model1} is the analogue of the EFT plot in Fig.~\ref{Fig:EFT1} (left), with the key difference that there is no constraint from $\cB(B\to K \nu\bar\nu)$, since this observable receives a vanishing contribution from the $U_1$ exchange, at tree level. In this simplified model, $\cB(B\to K \nu\bar\nu)$ receives tree-level NP contributions only via $Z^\prime$ exchange, as illustrated in Fig.~\ref{Fig:model1} (right). The $Z^\prime$ exchange also modifies $B_s$ mixing and the combination of these two effects allows us to determine both mass and mixing parameters of the $Z^\prime$, although with less precision with respect to the $U_1$ parameters, given NP effects induced by the $Z^\prime$ are smaller.

\begin{figure}[t]
\centering
\includegraphics[width = 8.5 cm]{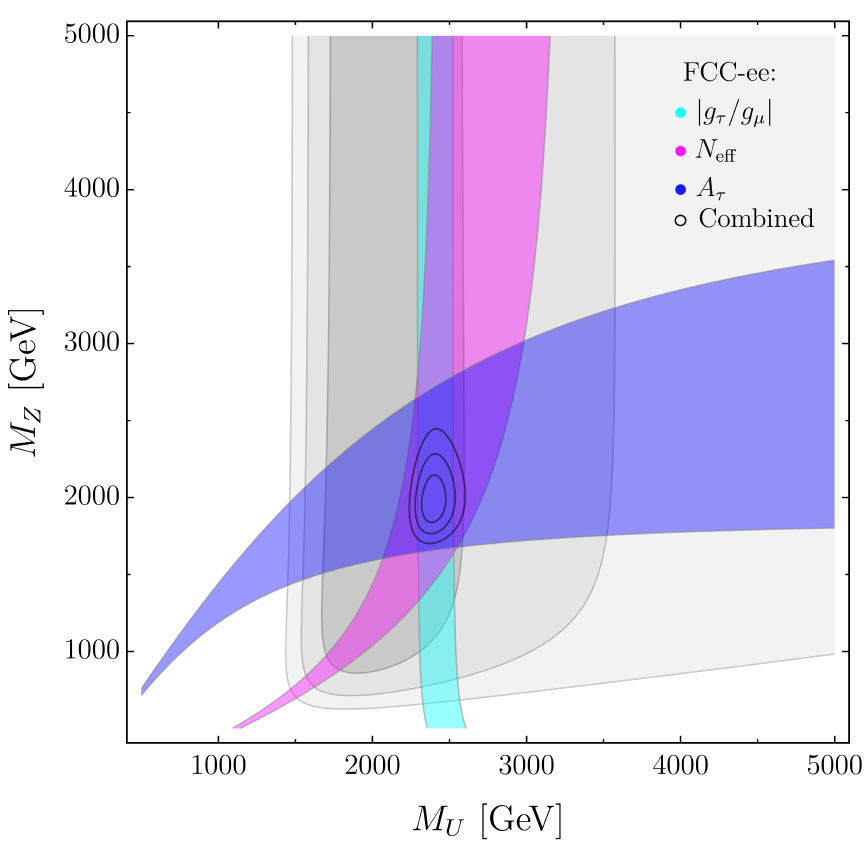}
\caption{Present and future constraints in the $\{ M_Z, M_U\}$ plane from selected electroweak and flavor observables in the $U_1$+$Z'$ model, assuming the NP benchmark in Eq.~(\ref{eq:benchM1}).
Notations as in Fig.~\ref{Fig:model1}. 
\label{Fig:model1EW}}
\end{figure}

In Fig.~\ref{Fig:model1EW} we analyze the impact of electroweak observables and LFU tests in $\tau$ decays, i.e.~the observables not sensitive to the flavor-mixing parameters ($\varepsilon_{U,Z}$), in constraining $M_U$ and $M_Z$. This plot should be compared with the one in Fig.~\ref{Fig:EFT3}, done in the EFT case, with the same set of observables. The difference, which is quite striking, highlights the difficulty in deducing the structure of the underlying theory solely from indirect observables. One of the great advantages of FCC-ee is the possibility to combine several independent observables, facilitating the reconstruction of the underlying theory in the (realistic) case where NP effects are described by several potentially independent parameters.

To better understand the role of electroweak observables in the $U_1$ + $Z'$ model, we present here their leading dependence from the Wilson coefficients in Eq.~(\ref{eq:Mod1-CU})--(\ref{eq:Mod1-CZ}). All the electroweak effects are generated beyond the tree level (having neglected a possible $Z^\prime$--$Z$ mixing), and we estimate them at the leading logarithmic accuracy from the running of the EFT coefficients. Keeping only $g_2$ and $y_t$ terms in the renormalization group equations (RGE) leads to the following approximate relations:
\begin{align}
    \delta A_\tau &\sim \cC_{H\ell}^{(1+3)} \approx (2g_2^2 \cC_{\ell q}^{(3)} + 6y_t^2 \cC_{\ell q}^{(1-3)}) L\,, \\
    \delta N_{\rm eff} &\sim \cC_{H\ell}^{(1-3)} \approx (-2g_2^2 \cC_{\ell q}^{(3)} + 6y_t^2 \cC_{\ell q}^{(1+3)}) L\,, 
\end{align}
where $L = (16\pi^2)^{-1} \log (m_Z/M_U)$ and all flavor indices are third-generation.
From the relations above (once we substitute the matching conditions) we observe that:
\begin{itemize}
    \item There is a cancellation between the $U_1$ and $Z'$ contributions to $A_\tau$ for $M_{Z}~= \sqrt{3/4}(y_t/g_2) M_U$.
 This is the origin of the fine-tuned region at small masses allowed by 
 $A_\tau$ (blue band) in Fig.~\ref{Fig:model1EW}.
 \item A similar cancellation happens within the $y_t^2$ terms in  $N_{\rm eff}$, leading to a vanishing constraint for $M_Z = M_U/2$.
 This is the origin of the fine-tuned region at small masses allowed by 
 $N_{\rm eff}$ (violet band) in Fig.~\ref{Fig:model1EW}.
\end{itemize}

\subsection{$S_1$}

Tree-level contributions to semileptonic operators 
can be also generated by scalar leptoquarks (see e.g.~\cite{Bauer:2015knc,Crivellin:2017zlb,Gherardi:2020det}). 
An interesting illustrative example is given by the case of a single heavy field $S_1 \sim(\overline{\mathbf{3}}, \mathbf{1})_{\frac{1}{3}}$ with the following Lagrangian
\begin{align}
    \mathcal{L}_{S_1} \supset iy_{L} S_1 \left(\bar q_L^{c\,3} \sigma_2 \ell_L^3 \right) +y_{R} S_1 \left(\bar u^{3c} e^{3}_R\right) + {\rm h.c.}\,
\end{align}
giving rise to the following tree-level matching conditions~\cite{Gherardi:2020det}:
\begin{align}
& {\cC_{l q}^{(1)[3333]}=-\frac{v^2}{2}\frac{y_{L}^{*} y_{L}}{4 M_S^2},} \qquad {\cC_{l q}^{(3)[3333]}=\frac{v^2}{2}\frac{y_{L}^{*} y_{L}}{4 M_S^2}} \\
& {\cC_{l e q u}^{(1)[3333]} = -\frac{v^2}{2}\frac{y_{R} y_{L}^{*}}{2 M_S^2},} \qquad {\cC_{l e q u}^{(3)[3333]}=\frac{v^2}{2}\frac{y_{R} y_{L}^{*}}{8 M_S^2}} \\
& {\cC_{e u}^{[3333]}=-\frac{v^2}{2}\frac{y_{R}^{*} y_{R}}{2 M_S^2}\,.}
\end{align}
Compared to the previous case, this simplified model allows us to analyze the impact of scalar flavor-changing operators and, most importantly, a possible correlation between flavor observables and precision measurements in Higgs decays.

\begin{table}[t]
    \centering
    \begin{tabular}{|c|c|c|c|}
        \hline
        Observable & SM   & FCC projection \\ \hline\hline
        $B_c \to \tau\nu$ & $(1.95 \pm 0.09)\times 10^{-2}$ &  $(1.99\pm 0.03)\times 10^{-2}$   \\ 
        $B\to K \nu\bar\nu$ & $(4.44 \pm 0.30)\times 10^{-6}$  & $(7.24 \pm 0.22)\times 10^{-6}$  \\
        $B\to K^* \nu\bar\nu$ & $(9.8 \pm 1.4)\times 10^{-6}$  & $(16.0 \pm 0.5)\times 10^{-6}$   \\ 
        $H \to \tau \bar \tau$ & $1$   & $1.13 \pm 0.01$  \\ \hline
         $\left|g_\tau/g_\mu\right|$ & 1.0   & 0.99958(70)   \\
        $A_\tau$ & $0.147$   & 0.14600(21) \\
        \hline
    \end{tabular}
    \caption{SM values and hypothetical projected values at FCC for the observables we consider in the $S_1$ model.\label{tab:S1}}
\end{table}

\begin{figure}[t]
\centering
\includegraphics[width = 7.3cm]{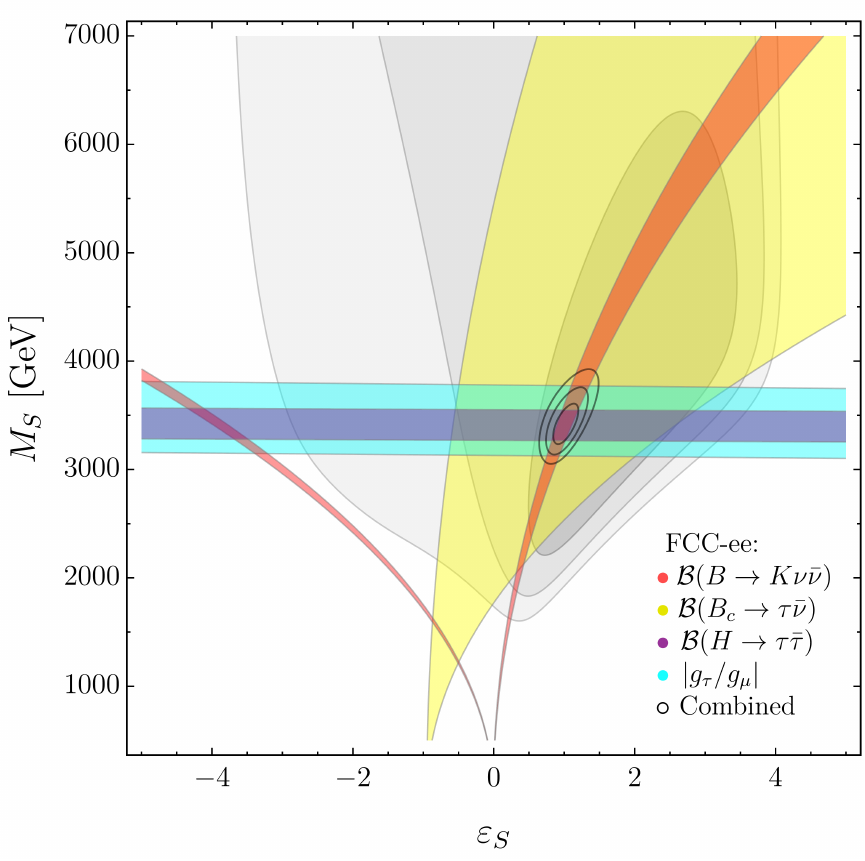}
\hskip 0.3 cm
\includegraphics[width = 7.3cm]{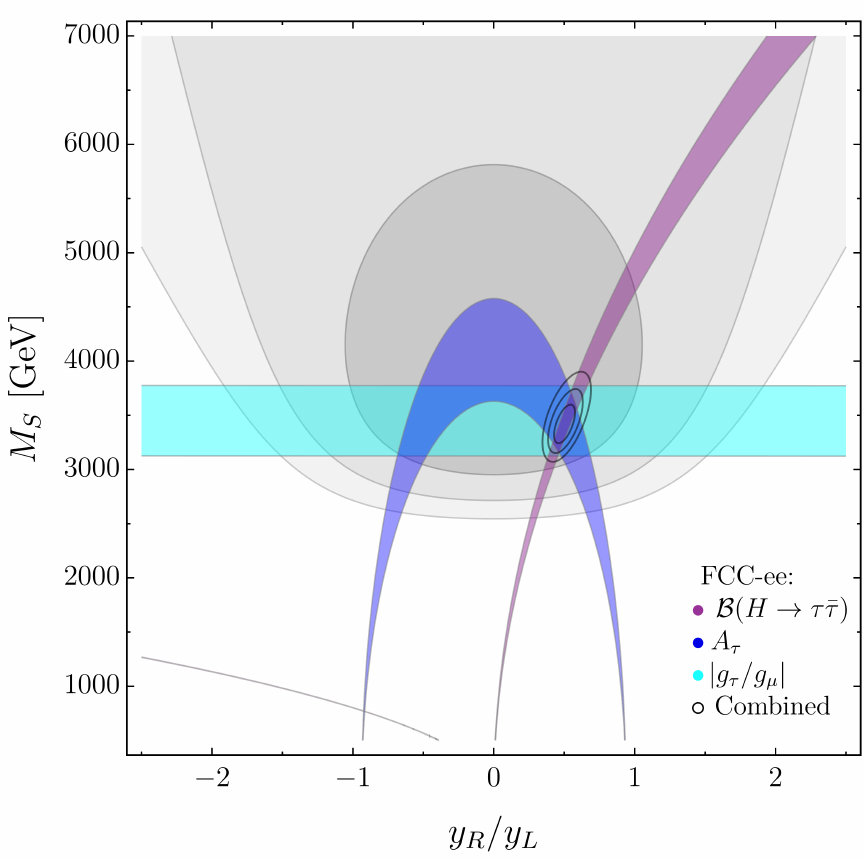}
\caption{Present and future constraints in the $\{ \varepsilon_S, M_S\}$ and 
 and  $\{ y_R/y_L, M_S \}$ planes in the $S_1$ model. Notations as in Fig.~\ref{Fig:model1}.
 Future constraints are obtained assuming the NP benchmark in Eq.~(\ref{eq:benchS1}). In all plots we 
 set $y_L=2$ (for different $y_L$ values, $M_S$ needs to be rescaled by $y_L/2$).
\label{Fig:model2} }
\end{figure}

Proceeding as above, we consider the following benchmark point:
\begin{equation}
    \{y_L= 2\, , \,\, M_S = 3.4\text{~TeV}\, ,  \,\, y_R =  1\, ,  \,\, \varepsilon_S= 1\, \} \,,
    \label{eq:benchS1}
\end{equation}
which implies $\cC_{\ell q}^{(3)}\approx 0.003$
(for third-generation fields).
Here $\varepsilon_S$ is the flavor-mixing parameter controlling off-diagonal couplings, defined in analogy to the $U_1$+$Z^\prime$ case. 
The projected values of the observables 
according to this benchmark point are reported in Table~\ref{tab:S1}.

In Fig.~\ref{Fig:model2} we illustrate the impact of both flavor, electroweak and Higgs observables in constraining the model.  In this case, there are only three parameters to which low-energy observables are sensitive to (we conventionally set $y_L=2$). As can be seen, an interesting feature of the model is a modified contribution to the $H\to \tau^+\tau^-$ width, controlled by the right-handed coupling ($y_R$). Precise measurements of Higgs couplings therefore play a key role in constraining the parameter space of the model, as shown in particular by the plot in Fig.~\ref{Fig:model2} (right). 
It is remarkable that the precision expected at FCC-ee is such that three very different sectors (flavor, electroweak and Higgs) contribute in an equally relevant manner in reconstructing the model.

\subsection{Vector-like fermions}

The last simplified model we consider is a framework where the SM is extended with the addition of heavy Vector-Like Fermions (VLFs). The existence of such fermions, whose chiralities transform identically under the SM gauge group, ensuring that their masses do not break electroweak symmetry, is a general expectation in many UV completions of the Standard Model (see e.g.~\cite{Ishiwata:2015cga,Bobeth:2016llm,Mann:2017wzh,Alves:2023ufm}).
Here we choose the following two specific VLF representations 
as an illustrative example,
\begin{align}
    D \sim (\mathbf{3},\mathbf{1},-1/3) \,, \qquad E \sim (\mathbf{1},\mathbf{1},-1) \,,
\end{align}
which can be coupled to the SM fields via 
\begin{align}
    \mathcal{L}_{\rm VLF} \supset \lambda_D \bar q_L^3 D_R H + \lambda_E \ell_L^3 E_R H + {\rm h.c.} 
\end{align}
\begin{table}[t]
    \centering
    \begin{tabular}{|c|c|c|c|}
        \hline
        Observable & SM   & FCC projection \\ \hline\hline 
        $B\to K \nu\bar\nu$ & $(4.44 \pm 0.30)\times 10^{-6}$  & $(4.93 \pm 0.15)\times 10^{-6}$  \\
        $B\to K^* \nu\bar\nu$ & $(9.8 \pm 1.4)\times 10^{-6}$  & $(10.9 \pm 0.3)\times 10^{-6}$   \\ \hline
         $\left|g_\tau/g_\mu\right|$ & 1.0   & 0.99960(71)   \\
        $A_\tau$ & $0.147$   & 0.14534(21) \\
        $R_b$ & $0.21581$   & 0.21549(6)  \\ 
        \hline
    \end{tabular}
    \caption{SM values and hypothetical projected values at FCC for the observables we consider in the VLFs model.\label{tab:VLF}}
\end{table}

This example is structurally different from the previous ones as well as from the EFT analysis, because no four-fermion operators are generated by integrating out the heavy fields at the tree level. The effect induced by the VLFs at the tree level is a mixing between fermion and Higgs currents that, after spontaneous symmetry breaking, leads to modified couplings for $W$ and $Z$ fields. The impact in flavor observables therefore arises by the induced flavor-changing couplings of the $Z$. On the other hand, the flavor-conserving part is tightly constrained by electroweak observables, leading again to an interesting interplay between the two different sets.

The relevant tree-level matching conditions read
\begin{align}
    &[C_{Hq}^{(1)}]_{33} = [C_{Hq}^{(3)}]_{33} = \frac{(\lambda_D)^2 v^2}{8 M_D^2}\,, \\
    &[C_{H\ell}^{(1)}]_{33} = [C_{H\ell}^{(3)}]_{33} = \frac{(\lambda_E)^2 v^2}{8 M_E^2}\,, \qquad\qquad  
    [C_{eH}]_{33} = - \frac{y_\tau v^2 \left(\lambda_E\right)^2}{4M_E^2}\,,
\end{align}
and we consider the following benchmark point:
\begin{equation}
    \{\lambda_D = \lambda_E = 0.5\, , \,\, M_E =3\text{~TeV}\,,  \,\, M_D = 3\text{~TeV}\, ,  \,\, \varepsilon_F = 1 \} \,.
   \label{eq:benchVLF}
\end{equation}
Here $\varepsilon_F$ is the flavor-mixing parameter controlling off-diagonal couplings, defined in analogy to the previous models. The projected values of the observables according to this benchmark point are reported in Table~\ref{tab:VLF}. 

\begin{figure}[t]
\centering
\includegraphics[width = 7.3cm]{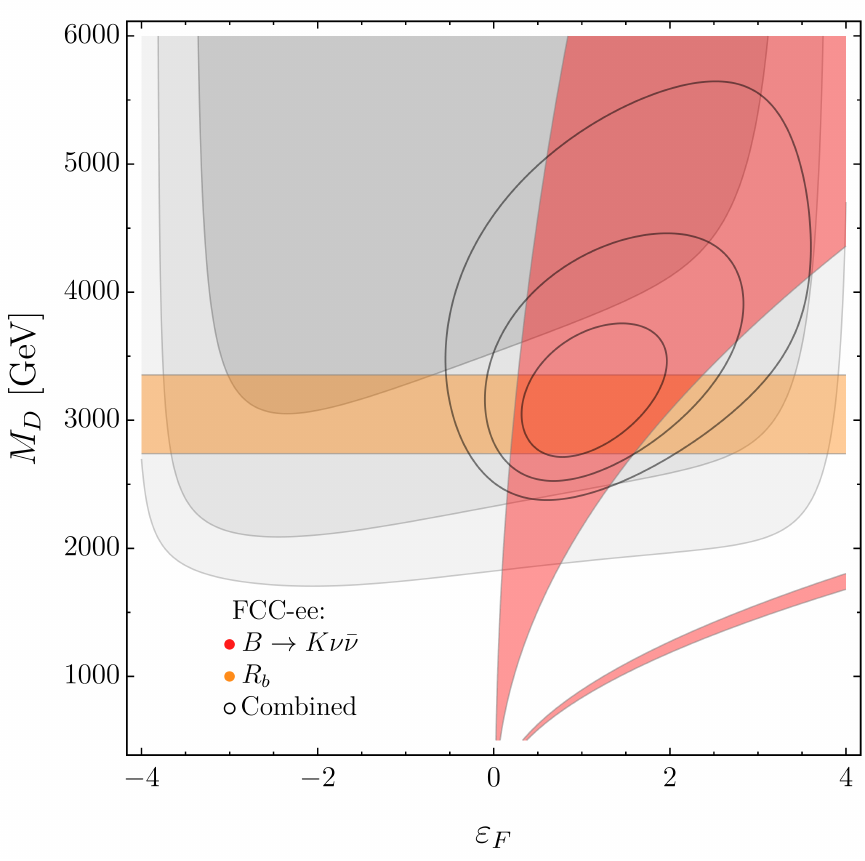}
\hskip 0.3 cm
\includegraphics[width = 7.3cm]{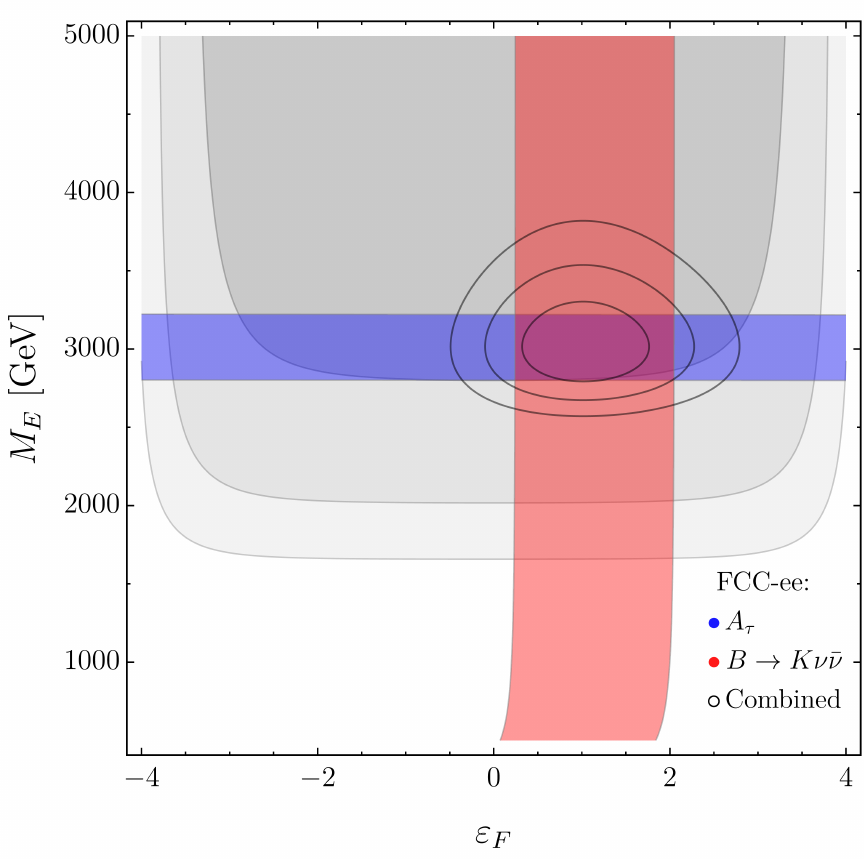}
\caption{Present and future constraints in the $\{ \varepsilon_F, M_D\}$ and  $\{ \varepsilon_F, M_E\}$ planes in the VLFs model. Notations as in Fig.~\ref{Fig:model1}.
 Future constraints are obtained assuming the NP benchmark in Eq.~(\ref{eq:benchVLF}).
\label{Fig:model3} In all plots, we set $\lambda_D = \lambda_E = 0.5$ (for different $\lambda_{D,E}$ values, $M_{D,E}$ need to be rescaled by $2\lambda_{D,E}$).}
\end{figure}

\begin{figure}[t]
\centering
\includegraphics[width = 8.5 cm]{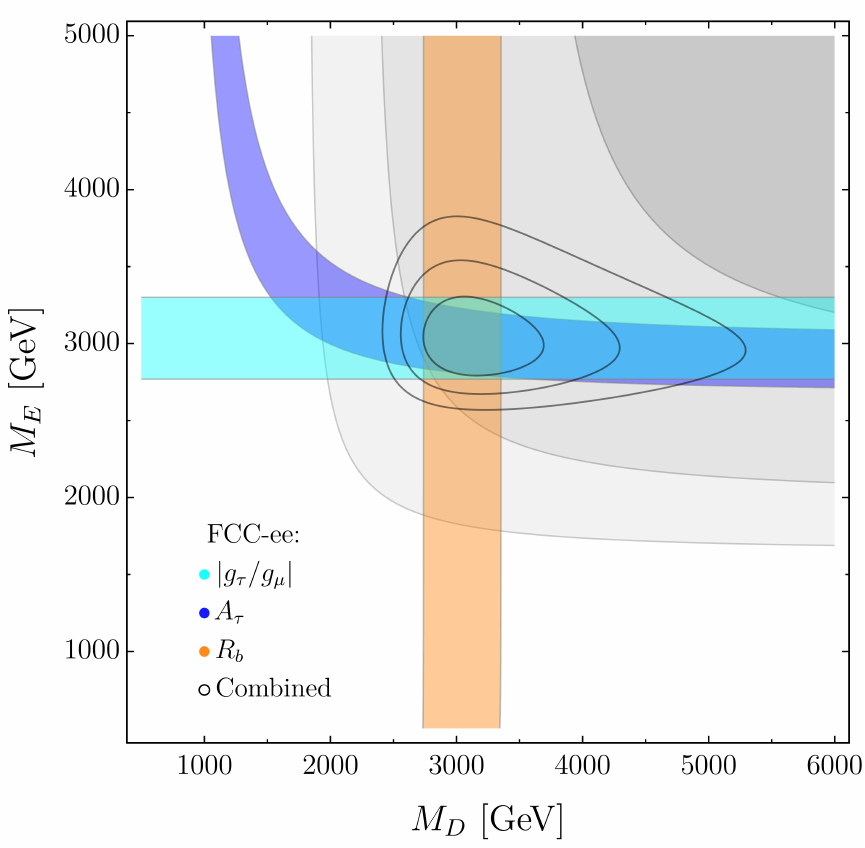}
\caption{Present and future constraints in the $\{ M_E, M_D\}$ plane in the VLFs model. Notations as in Fig.~\ref{Fig:model1}.
 Future constraints are obtained assuming the NP benchmark in Eq.~(\ref{eq:benchVLF}).
\label{Fig:model3EW}}
\end{figure}

There are only 3 effective parameters, as we conventionally set $\lambda_D = \lambda_E = 0.5$.
In Fig.~\ref{Fig:model3} and \ref{Fig:model3EW} we illustrate the impact of different observables in constraining the model. For example, in Fig.~\ref{Fig:model3} (left) we again observe a nice complementarity between electroweak observables, in this case represented by $R_b$, which is sensitive only to $M_D$, and the flavor observables, here $B\to K \nu\bar\nu$\footnote{Given that the effect in $b\to s$ transition proceeds only through a modified, flavor-changing $Z$ coupling, one expects a similar effect in $b\to s\tau\tau$ transitions. The projected precision however leads to a much larger region, completely overlapping with the one from $B\to K\nu\bar\nu$, which is why we choose to not show it here.}.
The combination leads to the determination of the parameter $\varepsilon_F$.
On the other hand, there is only a very mild dependence of the flavor observables on the mass of the vector-like lepton $E$.
It is interesting to notice that, while the two panels in Fig.~\ref{Fig:model3} seem to suggest a somewhat ``factorized" picture regarding electroweak observables, $A_\tau$ exhibits an interesting behavior in Fig.~\ref{Fig:model3EW}. The $M_D$ dependence
is a loop-suppressed effect, due to the running of $[\cC_{Hq}^{(1)}]_{33}$ into $\cC_{HD}$ proportional to the top Yukawa. $\cC_{HD}$ does not enter $A_\tau$ directly, but through the universal shift induced in the $\delta g$'s due to the choice of input scheme. Given that the $M_E$ contribution, on the other hand, is tree-level, this effect is relevant only for light $D$ masses. 

\subsection{Constraints on LFV couplings}

All the simplified models considered so far can be ``extended'' by including additional couplings, presumably suppressed, controlling flavor mixing in the lepton sector. 
This is a natural expectation in several realistic UV completions and, specifically, in models with 
non-universal gauge interactions. Flavor non-universality and flavor mixing in the lepton sector imply, in general, non-negligible rates for  Lepton Flavor Violating (LFV) processes with charged leptons, such as $\tau \to \mu \bar\mu \mu$ or $B_s\to \tau \bar \mu (\bar\mu \tau)$. Since such processes are strictly forbidden in the SM, the reconstruction of the model parameters factorizes: having fixed the lepton-flavor conserving parameters as illustrated in the previous sections, LFV processes can be used to bound or measure the effective LFV couplings.

As an illustrative case, we briefly comment here on LFV effects in the  $U_1+Z^\prime$ model. A first minimal extension is obtained by introducing the flavor mixing parameters $\varepsilon^\ell_{Z,U}$, describing $\ell^3_L$--$\ell^2_L$ mixing and defined in complete analogy with the quark-mixing parameters $\varepsilon_{Z,U}$. 
The strongest constraints  on  $\varepsilon^\ell_{U}$ are obtained by LFV semileptonic processes, such as 
$B_s\to \tau \bar \mu$ and $ B\to K \tau \bar \mu$. Assuming the benchmark point in~(\ref{eq:benchM1}), 
the Upgrade-II LHCb projections in Table~\ref{tab:exp} would allow us to set the bound $\varepsilon^\ell_{U}<0.2$
in absence of a signal.  Conversely, the FCC-ee projected sensitivity on $\tau \to \mu \bar\mu \mu$ would allow us to set the bound $\varepsilon^\ell_{Z}<0.1$ in absence of a signal. 

It is worth stressing that in both cases the effects, and the corresponding constraints, are strongly model-dependent. The $b\to s \tau\bar \mu$ rates mediated by the $U_1$ exchange  are proportional to $\varepsilon^2_U \times (\varepsilon^\ell_U)^2 $, hence the  bounds on $\varepsilon^\ell_U$ are largely affected by the precise value of  $\varepsilon_U$. In this minimal version of the model, the $Z^\prime$-mediated  $\tau \to \mu \bar\mu \mu$ rate is proportional to $(\varepsilon^\ell_Z)^6$; however, the dependence would change to  $(\varepsilon^\ell_Z)^2 \times (\xi_Z^\ell)^2$, if we also allow a 
flavor-conserving coupling of the $Z^\prime$ to light leptons (parameterized by  
$\xi_Z^\ell \ll 1$), as expected in many realistic models.\footnote{In models with flavor deconstruction, one expect $|\xi_Z^\ell| \approx g^2_{[12]}/g^2_{[3]}$, where $g_{[3]}$ ($g_{[12]}$) are the couplings of the gauge groups acting on third (light) families (see e.g.~\cite{Davighi:2023iks}).}  In view of these considerations, it is difficult to make definite statements 
about the impact of LFV observables; however, it is fair to state that for models with TeV-scale mediators coupled mainly to the third generation, as those analyzed in the previous sections, the potential bounds on 2-3 mixing in the left-handed lepton sector following from the projections in Table~\ref{tab:exp} are in the $O(10^{-1})$ range.

\section{NP Limits in the absence of deviations}
\label{sect:limits}

Moving away from the simplified scenarios analyzed in the previous sections we discuss here, in more generality, the expected sensitivities at FCC-ee to $U(2)^5$-symmetric SMEFT operators.
For this we do not inject a signal, but rather assume that the central values measured will be SM-like.
This analysis builds on the one presented in \cite{Allwicher:2023shc}, dividing the observables into three main categories:
\begin{itemize}
    \item \textbf{Collider}: This includes the tails of Drell-Yan distributions at the LHC, for which the likelihoods have been obtained with {\tt HighPT} \cite{Allwicher:2022mcg}, as well as four-lepton data from LEP-2 \cite{ALEPH:2013dgf}, and constraints on four-quark and QCD dipole operators from \cite{Hartland:2019bjb,Ethier:2021bye}.
    \item \textbf{Electroweak}: These are all the traditional pole observables, defined e.g. in \cite{Breso-Pla:2021qoe}, plus the signal strengths of $H\to f\bar f$, with $f = b,c,\tau$. Notice that, contrary to \cite{Allwicher:2023shc}, we choose to put LFU tests in $\tau$ decays in the next category (flavor), instead of assimilating them into the electroweak set.
    \item \textbf{Flavor}: In addition to the observables discussed in detail above, the analysis of the current flavor constraints is based on the larger set of observables defined in \cite{Allwicher:2023shc}. In particular, this includes also $D$- and $K$-meson mixing, $K\to \pi\nu\bar\nu$ decays, and the inclusive $B\to X_s\gamma$ rate. For all of these we do not include any future projections. Our results may be therefore considered as conservative.
\end{itemize}

\begin{figure}
    \centering
    \includegraphics[width=\linewidth]{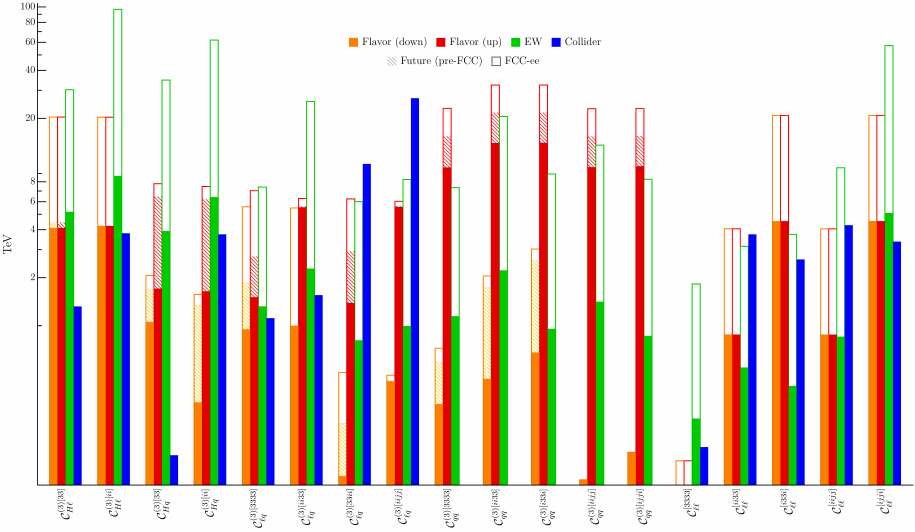}
    \caption{Bounds on a selection of $U(2)^5$-symmetric SMEFT operators from flavor, electroweak, and collider observables. The expected future sensitivities before the start of FCC are shown as hatched bars, while the empty bars represent the expected reach at FCC-ee. For all operators RG running from a reference scale of 1 TeV is taken into account, and all bounds are shown at 2$\sigma$.}
    \label{fig:barplot}
\end{figure}

The results of our analysis can be seen in Tables \ref{tab:U2boundsdown} and \ref{tab:U2boundsup}, and Figure \ref{fig:barplot}, where we report the most significant bounds.
The operators basis is the one defined in \cite{Faroughy:2020ina,Allwicher:2023shc}, i.e.~the Warsaw basis \cite{Grzadkowski:2010es} supplemented by an imposed $U(2)^5$ symmetry acting on the light generations.
In addition, given the misalignment of the mass eigenstates with respect to the weak interactions eigenstates for the left-handed quark doublet, it is also crucial to define the alignment within $q_L^3$ (cf. the discussion in Section \ref{sect:strategy}). 
We choose to present the bounds for both the up- and down-aligned cases.

Our analysis goes through the following steps: i) compute the observables and the likelihoods within the EFT at the appropriate energy scale, ii) RGE evolve within SMEFT/LEFT up to a reference scale of $\Lambda = 1$ TeV, using {\tt DsixTools} \cite{Fuentes-Martin:2020zaz}, and then iii) impose exact $U(2)^5$ symmetry at the high scale.
We then select only one operator at a time, and find the allowed interval (at 2$\sigma$) for the Wilson Coefficient, translating it into a scale.

From Figure~\ref{fig:barplot}, where we show a selection of the bounds, a clear complementarity emerges between the different sectors (flavor, EW, and collider), as already pointed out in Ref.~\cite{Allwicher:2023shc}.
The novelty here is the inclusion of the flavor projections, both for pre-FCC (i.e.~mostly LHCb  and Belle-II expectations), and FCC-ee ($Z$-pole run).
A first point to notice is that, even before the start of FCC-ee, the expected improvements in flavor-physics measurements  should lead to a major enhancement in the NP reach compared to the present bounds. This reach is further  enhanced by the Tera-Z flavor program, especially for semileptonic and pure leptonic operators.  
It is interesting to note the interplay between electroweak and flavor bounds, that in many cases both exceed the 10~TeV threshold.
Taking a closer look at the tables, one also notices a distinct pattern, which allows to identify the most relevant observables, and the corresponding expected reach in scale.
In the up-aligned case, for example, the highest scales are associated with $B_s$ mixing, reaching $\mathcal{O}(30)$ TeV. 
The bounds from $\tau$ LFU tests instead reach $\mathcal{O}(20)$ TeV, and are clearly independent of the alignment choice. The di-neutrino and 
di-tau transitions\footnote{Assuming no deviations from the SM, the projected relative uncertainties on the di-tau modes are increased by $75\%$ with respect to the entries in Table~\ref{tab:exp}.} also play an important role, more prominently in the up-aligned case, but still reaching the multi-TeV range also for down-alignment, due to RGE effects.

\begin{table}[]
    \centering
    \renewcommand{\arraystretch}{1.3}
    \resizebox{\textwidth}{!}{\input{downtable.txt}}
    \caption{Bounds on $U(2)^5$-symmetric operators in the down-aligned limit. We report only the operators for which the bound improves at FCC-ee with respect to current data and pre-FCC expectations, and for which $\Lambda \geq 500$ GeV. $\Lambda_{1,2}$ are the scales associated with the two endpoints of the $2\sigma$ intervals for the Wilson Coefficients, i.e. if $\mathcal{C} \in [\mathcal{C}_1,\mathcal{C}_2]$, then $\Lambda_i = v /\sqrt{2 \mathcal{C}_i } $, and we keep track of the sign. For future projections we only show the absolute bound, corresponding to $\min\{|\Lambda_1|,|\Lambda_2|\}$.}
    \label{tab:U2boundsdown}\label{tab:U2boundsdown}
\end{table}

\begin{table}[]
    \centering
    \renewcommand{\arraystretch}{1.3}
    \resizebox{\textwidth}{!}{\input{uptable.txt}}
    \caption{Bounds on $U(2)^5$-symmetric operators in the up-aligned limit. We report only the operators for which the bound improves at FCC-ee with respect to current data and pre-FCC expectations, and for which $\Lambda \geq 1$ TeV. Same notation as in Tab. \ref{tab:U2boundsdown}.}
    \label{tab:U2boundsup}
\end{table}

\section{Conclusions}
\label{sect:conclusions}

Flavor physics provides to this day one of the main possibilities we have for indirect discovery of NP.
With the upcoming data from Belle-II and LHCb during the planned High-Luminosity phase of the LHC, flavor physics will continue to offer valuable insights into the nature of BSM physics over the next two decades. In fact, given the projected statistical improvements, the sensitivity of flavor physics observables to new physics may surpass that of the searches performed at the high-energy frontier. However, it is also important to consider a longer-term perspective. Beyond this time frame, the so-called Tera-Z factories, next generation $e^+e^-$ colliders such as the FCC-ee or the CEPC, are being widely discussed in the community. With $\mathcal{O}(10^{12})$ or more produced $Z$ bosons, they provide an incredible   opportunity not only for electroweak precision physics, but for flavor as well, particularly for heavy flavors.
Indeed, the combination of high statistics in the relatively clean environment of a lepton collider allows for a significant step forward in precision.

In this work, we have analyzed the possible impact of flavor measurements at these future Tera-$Z$ factories, considering FCC-ee as concrete example, in probing extensions of the SM.
Motivated by patterns in current data, we focused on the scenario in which NP couples mainly to the third generation.
While being able to evade the strongest constraints from flavor through a $U(2)$-type symmetry protection, this possibility can also accommodate some of the tensions we currently observe in semileptonic $B$-meson transitions. Moreover, it aligns with the theoretically well-motivated scenario of TeV-scale NP aimed at minimizing the little hierarchy problem in the electroweak sector.
Approaching the problem from both an EFT and a simplified model perspective, we looked at scenarios with NP effects in semileptonic transitions, focusing in particular on the interplay between electroweak precision and flavor observables,  while considering current collider 
constraints.\footnote{In the scenarios we have considered, experiments at FCC-ee could provide additional constraints/evidences of NP from  
 precision measurements of  $\sigma( e^+e^-\to f\bar f)$ above the $Z$ pole~\cite{Greljo:2024ytg}. However, the analysis of these effects goes beyond the scope of this paper.}

Assuming a signal compatible with current data and projecting expected FCC-ee sensitivities, the picture that emerges clearly shows the incredible discovery potential of such a machine through precision measurements. It is worth noting that the NP scenarios we considered are far from being exhaustive; however, they provide a good illustration of realistic UV completions of the SM.  The set of observables considered is also far from being complete, but it is still sufficient to analyze the main correlations between flavor and electroweak physics.
In all the cases studied, the synergies between electroweak and flavor observables prove to be essential for identifying the model parameters, with particular attention to the $U(2)$-breaking terms.  
Complementarity and, in some cases, redundancy of different measurements is a key ingredient to validate a potential discovery via indirect measurements. 
While the chance of a discovery cannot be quantified, comparing the allowed parameter space of realistic NP models now and after the projected FCC-ee 
measurements provides a quantitative estimate of the potential reach of this machine.
Our analysis, despite being limited to a few representative cases, shows that this  potential is outstanding.

Taking a more pessimistic view, we have also examined the sensitivity to the NP scale in scenarios where no deviations from the SM are observed. This is done within the $U(2)^5$-symmetric SMEFT, and, once again, it highlights the potential reach of the FCC-ee, extending well beyond the $\mathcal{O}(10)$ TeV level.

\acknowledgments
We are very grateful to Claudia Cornella for interesting discussions and for providing invaluable help on the flavor likelihoods.
We also thank Marzia Bordone and Joe Davighi  for useful comments. 
The work of GI and MP is supported by the Swiss
National Science Foundation (SNF) under contract 200020 204428.
LA acknowledges funding from the Deutsche Forschungsgemeinschaft under Germany’s Excellence Strategy EXC 2121 “Quantum Universe” – 390833306, as well as from the grant 491245950.

\appendix
\section{Details on the construction of likelihoods}
\label{app:Flavobs}
The first step in constructing our likelihood is to express the theoretical prediction for each observable in terms of SMEFT WCs at a reference high scale, which we take to be $\Lambda_{\rm NP}=1$ TeV. For observables defined at a scale below the EW scale, this requires evolving the relevant LEFT WCs up to the EW scale, matching to SMEFT, and then running in the SMEFT up to $\Lambda_{\rm NP}$. Running effects within the LEFT are particularly significant for the scalar and tensor operators, which enter observables in the charged-current $b \rightarrow c(u) \ell \nu$ and neutral-current $b \rightarrow s \ell \ell$ transitions. The SMEFT-to-LEFT matching is done at tree level.

Once all observables are expressed in terms of SMEFT WCs at $\Lambda_{\rm NP}$, we construct the likelihood as:
\begin{equation}
\chi^2\left(\vec{\mathcal{C}}\left(\Lambda_{\mathrm{NP}}\right)\right)=\sum_{i j}\left[\mathcal{O}_i^{\text {th }}\left(\vec{\mathcal{C}}\left(\Lambda_{\mathrm{NP}}\right)\right)-\mathcal{O}_i^{\exp }\right] \sigma_{i j}^{-2}\left[\mathcal{O}_j^{\text {th }}\left(\vec{\mathcal{C}}\left(\Lambda_{\mathrm{NP}}\right)\right)-\mathcal{O}_j^{\exp }\right]
\end{equation}
where $\mathcal{O}_i^{\text {th }}\left(\vec{\mathcal{C}}\left(\Lambda_{\mathrm{NP}}\right)\right)$ is the theory prediction for the observable $\mathcal{O}_i$ written in terms of the high-scale WCs, $\mathcal{O}_i^{\exp }$ is the corresponding experimental value and $\sigma_{i j}$ is the covariance matrix for $\mathcal{O}_i$ and $\mathcal{O}_j$. The numerical values of the current experimental measurements and their associated uncertainties\footnote{For measurements where only an upper bound (68\% C.L.) is available, the upper bound itself is taken as the experimental uncertainty, taking zero as the central value.}, along with the correlations and the projected sensitivities at FCC-ee, are presented in Tables \ref{tab:exp} and \ref{tab:ewprojections}.
\subsection{$\tau$ LFU tests}
The observable are defined as in \cite{Allwicher:2021ndi}:

\begin{align}
    \left|\frac{g_{\tau}}{g_{\mu}}\right|^2 \equiv \frac{\Gamma(\tau \rightarrow e \nu \bar{\nu})}{\Gamma(\mu \rightarrow e \nu \bar{\nu})}\left[\frac{\Gamma_{\mathrm{SM}}(\tau \rightarrow e \nu \bar{\nu})}{\Gamma_{\mathrm{SM}}(\mu \rightarrow e \nu \bar{\nu})}\right]^{-1} \,, \\
    \left|\frac{g_{\tau}}{g_{e}}\right|^2 \equiv \frac{\Gamma(\tau \rightarrow \mu \nu \bar{\nu})}{\Gamma(\mu \rightarrow e \nu \bar{\nu})}\left[\frac{\Gamma_{\mathrm{SM}}(\tau \rightarrow \mu \nu \bar{\nu})}{\Gamma_{\mathrm{SM}}(\mu \rightarrow e \nu \bar{\nu})}\right]^{-1} \,,
\end{align}
with the correlation $\rho=0.51$. With our conventions, we get:
\begin{align}
& \left|\frac{g_\tau}{g_\mu}\right| \approx 1+2\operatorname{Re}\left(-\mathcal{C}_{H \ell}^{(3)[33]} +\mathcal{C}_{\ell \ell}^{[1331]}+\mathcal{C}_{H \ell}^{(3)[22]} -\mathcal{C}_{\ell \ell}^{[1221]}\right)(\mu = m_Z) \\
& \left|\frac{g_\tau}{g_e}\right| \approx 1+2\operatorname{Re}\left(-\mathcal{C}_{H \ell}^{(3)[33]} +\mathcal{C}_{\ell \ell}^{[2332]}+\mathcal{C}_{H \ell}^{(3)[11]} -\mathcal{C}_{\ell \ell}^{[1221]}\right)(\mu = m_Z) \, .
\end{align}
The above expressions are given in the \textit{redundant} basis. To convert to the non-redundant one sets $\cC_{\ell\ell}^{[jiij]} = \cC_{\ell\ell}^{[ijji]}$ for $i\leq j$.

\subsection{$b\to c \ell \nu$ transitions}
The expressions for $R_D$ and $R_{D^*}$ at $\mu=m_b$ are taken from \cite{Cornella:2021sby,Aebischer:2022oqe}, and then evolved to $\mu=\mu_{EW}$. The running of $C_{V_L}$ is neglected and we only consider the running of the scalar operator $C_{S_X}^i(m_b)=1.46\,C_{S_X}^i(\mu_{EW})$. The contribution of LFV operators is also ignored.
\begin{align}
R_D & =R_D^{\mathrm{SM}}\left[\left|1+C_{V_L}\right|^2+2.19 \operatorname{Re}\left\{\left(1+C_{V_L}\right) C_{S_R}^{*}\right\}+2.195\left|C_{S_R}\right|^2\right] \quad \mu=\mu_{EW}\\
R_{D^*} & =R_{D^*}^{\mathrm{SM}}\left[\left|1+C_{V_L}\right|^2+0.175 \operatorname{Re}\left\{\left(1+C_{V_L}\right) C_{S_R}^{*}\right\}+0.085\left|C_{S_R}\right|^2\right] \quad \mu=\mu_{EW}
\end{align}
with correlation $\rho=-0.39$. The $C_i$ are given by \cite{Allwicher:2022gkm}
\begin{align}
C_{V_L} & = 2 \sum_i \frac{V_{c i}}{V_{cb}}\left(\mathcal{C}_{l q}^{(3) [33 i b] }-\mathcal{C}_{H q}^{(3) [i3]} -\delta_{i 3}\mathcal{C}_{H l}^{(3)[33] } \right) \, ,\\
C_{S_R} & =  \sum_i \frac{V_{c i} } {V_{cb}}\left(\mathcal{C}^{[333i]}_{\text {ledq }}\right)^*\, .
\end{align}
Similarly, the expression for the $B_c \to \tau \nu$ branching fraction at $\mu=\mu_{EW}$ is given by
\begin{align}
    \mathcal{B}(B_c \to \tau\nu)= \mathcal{B}(B_c\to \tau\nu )^{\rm SM} × \left| 1 + C_{V_L}
    + 
   1.46 C_{S_R} 
    \frac{m_{B_c}^2}{m_\tau (m_b + m_c)} \right|^2 \quad \mu=\mu_{EW}\,.
\end{align}

\subsection{$b\to s \nu \bar\nu$ transitions} 

We define
\begin{align}
\mathcal{B}(B\to {K^{(\ast)}}{\nu\nu}) &= \mathcal{B}(B\to {K^{(\ast)}}{\nu\nu})\Big{|}_\mathrm{SM} \, \big{(}1+\delta \mathcal{B}_{K^{(\ast)}}^{\nu\nu}\big{)}\,,
\end{align}
with $C_L^\mathrm{SM}= -6.32(7)$
, and
\begin{align}
\delta \mathcal{B}_{K^{(\ast)}}^{\nu\bar{\nu}} &=  \sum_{i}\dfrac{2\mathrm{Re}[C_L^\mathrm{SM}\,(\delta C_{L}^{ii}+\delta C_{R}^{ii})]}{3|C_{L}^\mathrm{SM}|^2}
+\sum_{i,j}\dfrac{|\delta C_{L}^{ij}+\delta C_{R}^{ij}|^2}{3|C_L^\mathrm{SM}|^2}+\\
&- \eta_{K^{(\ast)}}\sum_{i,j} \dfrac{\mathrm{Re}[\delta C_R^{ij}(C_{L}^\mathrm{SM}\delta_{ij}+\delta C_{L}^{ij})]}{3|C_{L}^\mathrm{SM}|^2}\,,
\end{align}
with $\eta_K=0$, and $\eta_{K^\ast}=3.33(7)$.
Running effects are absent here, and the $\delta C$ are given in terms of SMEFT operators at the electroweak scale as
\begin{align}
    \delta C_L^{ij} &= - \frac{2\pi}{\alpha}\frac{1}{V_{tb}V_{ts}^*} \left(\mathcal{C}_{\ell q}^{(1)[ij23]}-\mathcal{C}_{\ell q}^{(3)[ij23]}+\mathcal{C}_{H q}^{(1)[23]}+\mathcal{C}_{H q}^{(3)[23]}\right)\,, \\
    \delta C_R^{ij} &= - \frac{2\pi}{\alpha}\frac{1}{V_{tb}V_{ts}^*} \left(\mathcal{C}_{\ell d}^{[ij23]}+\mathcal{C}_{H d}^{[23]}\right)\,.
\end{align}

\subsection{$b\to s \tau \tau$ transitions}
The observables at $\mu=m_b$ are taken from \cite{Cornella:2021sby} and then evolved to $\mu=m_Z$, neglecting the running for $C_9$ and $C_{10}$, while for the scalar and pseudoscalar operators we find $C_{S(P)}^{\alpha\beta} (m_b) = 1.38\, C_{S(P)}^{\alpha\beta}(m_Z)$ (using {\tt DsixTools}). The expressions at $\mu=m_Z$ are given by:
\begin{align}
\nonumber
\mathcal{B}\left(B \rightarrow K \tau^{+} \tau^{-}\right) &= 10^{-9}\left(2.2\left|\mathcal{C}_9^{33}\right|^2+6.0\left|\mathcal{C}_{10}^{33}\right|^2+15.8\left|\mathcal{C}_S^{33}\right|^2+16.95\left|\mathcal{C}_P^{33}\right|^2\right. \\
& \left.+6.62 \operatorname{Re}\left\{\mathcal{C}_S^{33} \mathcal{C}_9^{{33} *}\right\}+8.14 \operatorname{Re}\left\{\mathcal{C}_P^{33} \mathcal{C}_{10}^{{33} *}\right\}\right)\,, \\
\nonumber
\mathcal{B}\left(B_s \rightarrow \tau^{+} \tau^{-}\right)&=\mathcal{B}\left(B_s \rightarrow \tau^{+} \tau^{-}\right)_{\mathrm{SM}}\Bigg\{  \left|1+\frac{\mathcal{C}_{10, \mathrm{NP}}^{33}}{\mathcal{C}_{10, \mathrm{SM}}}+1.38 \frac{\mathcal{C}_P^{33}}{C_{10, \mathrm{SM}}} \frac{m_{B_s}^2}{2 m_{\tau}\left(m_b+m_s\right)}\right|^2 \\
& \left.+\left(1-\frac{4 m_{\tau}^2}{m_{B_s}^2}\right)\left|1.38\frac{\mathcal{C}_S^{33}}{\mathcal{C}_{10, \mathrm{SM}}} \frac{m_{B_s}^2}{2 m_{\tau}\left(m_b+m_s\right)}\right|^2\right\}\, ,
\end{align}
where we write
\begin{equation}
    \mathcal{C}_k^{\alpha \beta}=\mathcal{C}_{k, \mathrm{SM}}^{\alpha \beta}+\mathcal{C}_{k, \mathrm{NP}}^{i \alpha \beta}
\end{equation}
and
\begin{align}
C_{9,\mathrm{SM}}=4.114 \quad C_{10,\mathrm{SM}}=-4.193 \quad C_{S,\mathrm{SM}} = 0 \quad C_{P,\mathrm{SM}} = 0\,.
\end{align}

\noindent
In terms of SMEFT operators, the $C_k$ are given by
\begin{align}
    C_{9,\rm NP}^{\alpha\beta} &= -\frac{2\pi}{\alpha V_{tb}V_{ts}^*}\left[ \cC_{qe}^{[23\alpha\beta]} + \cC_{\ell q}^{(1)[\alpha\beta 23]} + \cC_{\ell q}^{(3)[\alpha\beta 23]} + \delta^{\alpha\beta}(4s^2_W - 1)\left(\cC_{Hq}^{(1)[23]} + \cC_{Hq}^{(3)[23]}\right) \right]\,, \\ 
    C_{10,\rm NP}^{\alpha\beta} &= -\frac{2\pi}{\alpha V_{tb}V_{ts}^*}\left[ \cC_{qe}^{[23\alpha\beta]} - \cC_{\ell q}^{(1)[\alpha\beta 23]} - \cC_{\ell q}^{(3)[\alpha\beta 23]} + \delta^{\alpha\beta}\left(\cC_{Hq}^{(1)[23]} + \cC_{Hq}^{(3)[23]}\right) \right]\,, \\
    C_{S,\rm NP}^{\alpha\beta} &= -C_{P,\rm NP}^{\alpha\beta} = -\frac{2\pi}{\alpha V_{tb}V_{ts}^*} \cC_{\ell edq}^{[\beta\alpha 32]*}\,,
\end{align}
where $s_W$ is the sine of the Weinberg angle.

\subsection{$\Delta F = 2$}
We define the observable $\Delta M_{B_s}/\Delta M^{\rm SM}_{B_s}$ as
\begin{equation}
    \frac{\Delta M_{B_s}}{\Delta M^{\rm SM}_{B_s}} \equiv 1 + \frac{\mathcal{C}^{\rm NP}_{B_s}}{\mathcal{C}^{\rm SM}_{B_s}}\,,
\end{equation}
where the SM contribution at $\mu=\mu_{EW}$ is given by
\begin{equation}
\mathcal{C}^{\rm SM}_{B_s}=\frac{G_F^2 M_W^2}{4\pi^2}\left(V_{t b} V_{t s}^*\right)^2 S_0\left(\frac{M_t^2}{M_W^2}\right) \approx 8.3 \times 10^{-11} \text{~GeV}^{-2}\,.
\end{equation}
In the down-basis and using our SMEFT convention, the matching \cite{Allwicher:2023shc} is given by
\begin{equation}
    \mathcal{C}^{\rm NP}_{B_s}=-\left(\frac{2}{v^2}\right)\left(\mathcal{C}_{q q}^{(1)[2323]}+\mathcal{C}_{q q}^{(3)[2323]}\right)\,.
\end{equation}

\subsection{Higgs decays}
The Higgs signal strength are given by \cite{Allwicher:2023shc}
\begin{equation}
\mu_{i i}=\frac{\mathcal{B}\left(h \rightarrow f_i \bar{f}_i\right)}{\mathcal{B}\left(h \rightarrow f_i \bar{f}_i\right)_{\mathrm{SM}}}=\left|1+2\frac{v}{\sqrt{2} m_i^f}\left[\mathcal{C}_{f H}\right]_{i i}\right|^2 \, .
\end{equation}

\subsection{Electroweak observables}
\label{app:EWobs}
The expressions for electroweak observables are taken from the appendices in \cite{Allwicher:2023aql} and adapted to our SMEFT convention. 

\bibliographystyle{JHEP}
\bibliography{refs}

\end{document}

%% file: downtable.txt
\begin{tabular}{c|cccc|cc|cc}
\multirow{2}{*}{coeff.} & \multicolumn{4}{|c|}{Current data} & \multicolumn{2}{|c|}{Future (pre-FCC)} & \multicolumn{2}{|c}{FCC proj.}\\
  & $\Lambda_{1}$ [TeV] & Obs. & $\Lambda_{2}$ [TeV] & Obs. & $\Lambda$ [TeV] & Obs. & $\Lambda$ [TeV] & Obs.\\ \hline\hline
$\mathcal{C}_{\ell \ell }^{[ijji]}$ & -4.5 & $\tau$ LFU & 4.5 & $\tau$ LFU & 4.5 & $\tau$ LFU & 20.8 & $\tau$ LFU \\
$\mathcal{C}_{\ell \ell }^{[i33i]}$ & -4.5 & $\tau$ LFU & 4.5 & $\tau$ LFU & 4.5 & $\tau$ LFU & 20.8 & $\tau$ LFU \\
$\mathcal{C}_{H\ell }^{(3)[33]}$ & -4.8 & $\tau$ LFU & 4.1 & $\tau$ LFU & 4.4 & $\tau$ LFU & 20.4 & $\tau$ LFU \\
$\mathcal{C}_{H\ell }^{(3)[ii]}$ & -4.2 & $\tau$ LFU & 4.7 & $\tau$ LFU & 4.2 & $\tau$ LFU & 20.3 & $\tau$ LFU \\
$\mathcal{C}_{\ell q}^{(3)[3333]}$ & -2.2 & $R_D$/$R_{D^*}$ & -0.9 & $\tau$ LFU & 1.9 & $R_D$/$R_{D^*}$ & 5.6 & $\tau$ LFU \\
$\mathcal{C}_{\ell q}^{(3)[ii33]}$ & -1.8 & $\tau$ LFU & 1. & $\tau$ LFU & 1. & $\tau$ LFU & 5.5 & $\tau$ LFU \\
$\mathcal{C}_{\ell \ell }^{[ii33]}$ & -0.9 & $\tau$ LFU & 0.9 & $\tau$ LFU & 0.9 & $\tau$ LFU & 4.1 & $\tau$ LFU \\
$\mathcal{C}_{\ell \ell }^{[iijj]}$ & -0.9 & $\tau$ LFU & 0.9 & $\tau$ LFU & 0.9 & $\tau$ LFU & 4.1 & $\tau$ LFU \\
$\mathcal{C}_{\ell edq}^{[3333]}$ & -1.5 & $R_D$/$R_{D^*}$ & -0.6 & $R_D$/$R_{D^*}$ & 1.1 & $R_D$/$R_{D^*}$ & 3.6 & $B_c \to \tau\nu$ \\
$\mathcal{C}_{qq}^{(3)[i33i]}$ & 0.7 & $B\to K\nu\bar\nu$ & 0.7 & $B\to K\nu\bar\nu$ & 2.6 & $B\to K\nu\bar\nu$ & 3. & $B\to K\nu\bar\nu$ \\
$\mathcal{C}_{qq}^{(1)[i33i]}$ & -0.7 & $B\to K\nu\bar\nu$ & -0.6 & $B\to K\nu\bar\nu$ & 2.4 & $B\to K\nu\bar\nu$ & 2.8 & $B\to K\nu\bar\nu$ \\
$\mathcal{C}_{Hq}^{(3)[33]}$ & 1.1 & $B\to K\nu\bar\nu$ & 2.3 & $R_D$/$R_{D^*}$ & 1.7 & $B\to K\nu\bar\nu$ & 2.1 & $B\to K\nu\bar\nu$ \\
$\mathcal{C}_{qq}^{(3)[ii33]}$ & -0.5 & $B\to K\nu\bar\nu$ & -0.5 & $B\to K\nu\bar\nu$ & 1.7 & $B\to K\nu\bar\nu$ & 2.1 & $B\to K\nu\bar\nu$ \\
$\mathcal{C}_{Hq}^{(3)[ii]}$ & 0.3 & $B\to K\nu\bar\nu$ & 0.3 & $B\to K\nu\bar\nu$ & 1.3 & $B\to K\nu\bar\nu$ & 1.6 & $B\to K\nu\bar\nu$ \\
$\mathcal{C}_{Hu}^{[33]}$ & 0.8 & $B\to K\nu\bar\nu$ & 1.2 & $B\to K^{*}\nu\bar\nu$ & 0.6 & $B\to K\nu\bar\nu$ & 0.7 & $B\to K\nu\bar\nu$ \\
$\mathcal{C}_{qq}^{(3)[3333]}$ & 0.3 & $B\to K\nu\bar\nu$ & 0.6 & $R_D$/$R_{D^*}$ & 0.6 & $R_D$/$R_{D^*}$ & 0.7 & $B\to K\nu\bar\nu$ \\
$\mathcal{C}_{qu}^{(1)[ii33]}$ & -0.3 & $B_s$ mixing & 0.5 & $B_s$ mixing & 0.5 & $B_s$ mixing & 0.7 & $B_s$ mixing \\
$\mathcal{C}_{qu}^{(8)[ii33]}$ & -0.3 & $B_s$ mixing & 0.5 & $B_s$ mixing & 0.4 & $B_s$ mixing & 0.7 & $B_s$ mixing \\
$\mathcal{C}_{H\ell }^{(1)[33]}$ & -1.1 & $B\to K^{*}\nu\bar\nu$ & 0.2 & $B\to K\nu\bar\nu$ & 0.3 & $B\to K\nu\bar\nu$ & 0.7 & $B_s\to \tau\tau$ \\
$\mathcal{C}_{\ell u}^{[3333]}$ & -0.2 & $B\to K\nu\bar\nu$ & 1.5 & $B\to K^{*}\nu\bar\nu$ & 0.3 & $B\to K\nu\bar\nu$ & 0.7 & $B_s\to \tau\tau$ \\
$\mathcal{C}_{\ell q}^{(1)[3333]}$ & 0.1 & $B\to K\nu\bar\nu$ & 0.2 & $B\to K\nu\bar\nu$ & 0.3 & $B\to K\nu\bar\nu$ & 0.6 & $B_s\to \tau\tau$ \\
$\mathcal{C}_{qu}^{(1)[3333]}$ & -0.3 & $B_s$ mixing & 0.7 & $B_s$ mixing & 0.4 & $B\to K\nu\bar\nu$ & 0.5 & $B_s$ mixing \\
$\mathcal{C}_{\ell q}^{(1)[ii33]}$ & 0.4 & $\tau$ LFU & 0.8 & $B\to K\nu\bar\nu$ & 0.4 & $B\to K\nu\bar\nu$ & 0.5 & $\tau$ LFU \\
$\mathcal{C}_{\ell q}^{(3)[33ii]}$ & -0.2 & $B\to K\nu\bar\nu$ & -0.1 & $B\to K\nu\bar\nu$ & 0.2 & $B\to K\nu\bar\nu$ & 0.5 & $B_s\to \tau\tau$ \\
$\mathcal{C}_{qu}^{(8)[3333]}$ & -0.2 & $B_s$ mixing & 0.4 & $B_s$ mixing & 0.3 & $B_s$ mixing & 0.5 & $B_s$ mixing \\
$\mathcal{C}_{\ell q}^{(1)[33ii]}$ & 0.1 & $B\to K\nu\bar\nu$ & 0.6 & $R_D$/$R_{D^*}$ & 0.2 & $B\to K\nu\bar\nu$ & 0.5 & $B_s\to \tau\tau$ \\
 \hline
\end{tabular}

%% file: uptable.txt
\begin{tabular}{c|cccc|cc|cc}
\multirow{2}{*}{coeff.} & \multicolumn{4}{|c|}{Current data} & \multicolumn{2}{|c|}{Future (pre-FCC)} & \multicolumn{2}{|c}{FCC proj.}\\
  & $\Lambda_{1}$ [TeV] & Obs. & $\Lambda_{2}$ [TeV] & Obs. & $\Lambda$ [TeV] & Obs. & $\Lambda$ [TeV] & Obs.\\ \hline\hline
$\mathcal{C}_{qq}^{(1)[i33i]}$ & -14.2 & $B_s$ mixing & 14.2 & $B_s$ mixing & 21.8 & $B_s$ mixing & 32.5 & $B_s$ mixing \\
$\mathcal{C}_{qq}^{(1)[ii33]}$ & -14.4 & $B_s$ mixing & 14. & $B_s$ mixing & 21.7 & $B_s$ mixing & 32.4 & $B_s$ mixing \\
$\mathcal{C}_{qq}^{(3)[i33i]}$ & -14.3 & $B_s$ mixing & 14. & $B_s$ mixing & 21.7 & $B_s$ mixing & 32.4 & $B_s$ mixing \\
$\mathcal{C}_{qq}^{(3)[ii33]}$ & -14. & $B_s$ mixing & 14.4 & $B_s$ mixing & 21.7 & $B_s$ mixing & 32.4 & $B_s$ mixing \\
$\mathcal{C}_{qq}^{(1)[ijji]}$ & -10.1 & $B_s$ mixing & 10.1 & $B_s$ mixing & 15.5 & $B_s$ mixing & 23.1 & $B_s$ mixing \\
$\mathcal{C}_{qq}^{(1)[3333]}$ & -9.8 & $B_s$ mixing & 10.5 & $B_s$ mixing & 15.4 & $B_s$ mixing & 23.1 & $B_s$ mixing \\
$\mathcal{C}_{qq}^{(1)[iijj]}$ & -10.2 & $B_s$ mixing & 10. & $B_s$ mixing & 15.5 & $B_s$ mixing & 23.1 & $B_s$ mixing \\
$\mathcal{C}_{qq}^{(3)[3333]}$ & -10.4 & $B_s$ mixing & 9.8 & $B_s$ mixing & 15.4 & $B_s$ mixing & 23.1 & $B_s$ mixing \\
$\mathcal{C}_{qq}^{(3)[ijji]}$ & -10. & $B_s$ mixing & 10.1 & $B_s$ mixing & 15.5 & $B_s$ mixing & 23.1 & $B_s$ mixing \\
$\mathcal{C}_{qq}^{(3)[iijj]}$ & -9.9 & $B_s$ mixing & 10.3 & $B_s$ mixing & 15.4 & $B_s$ mixing & 23. & $B_s$ mixing \\
$\mathcal{C}_{\ell \ell }^{[ijji]}$ & -4.5 & $\tau$ LFU & 4.5 & $\tau$ LFU & 4.5 & $\tau$ LFU & 20.8 & $\tau$ LFU \\
$\mathcal{C}_{\ell \ell }^{[i33i]}$ & -4.5 & $\tau$ LFU & 4.5 & $\tau$ LFU & 4.5 & $\tau$ LFU & 20.8 & $\tau$ LFU \\
$\mathcal{C}_{H\ell }^{(3)[33]}$ & -4.8 & $\tau$ LFU & 4.1 & $\tau$ LFU & 4.4 & $\tau$ LFU & 20.4 & $\tau$ LFU \\
$\mathcal{C}_{H\ell }^{(3)[ii]}$ & -4.2 & $\tau$ LFU & 4.7 & $\tau$ LFU & 4.2 & $\tau$ LFU & 20.3 & $\tau$ LFU \\
$\mathcal{C}_{\ell edq}^{[3333]}$ & -0.5 & $B_s\to \tau\tau$ & 1. & $B_s\to \tau\tau$ & 0.8 & $B_s\to \tau\tau$ & 9.2 & $B_s\to \tau\tau$ \\
$\mathcal{C}_{Hq}^{(3)[33]}$ & 1.7 & $B\to K\nu\bar\nu$ & 1.8 & $B\to K\nu\bar\nu$ & 6.5 & $B\to K\nu\bar\nu$ & 7.8 & $B\to K\nu\bar\nu$ \\
$\mathcal{C}_{Hq}^{(3)[ii]}$ & -1.7 & $B\to K\nu\bar\nu$ & -1.6 & $B\to K\nu\bar\nu$ & 6.2 & $B\to K\nu\bar\nu$ & 7.5 & $B\to K\nu\bar\nu$ \\
$\mathcal{C}_{Hq}^{(1)[ii]}$ & -1.7 & $B\to K\nu\bar\nu$ & -1.6 & $B\to K\nu\bar\nu$ & 6.1 & $B\to K\nu\bar\nu$ & 7.4 & $B\to K\nu\bar\nu$ \\
$\mathcal{C}_{Hq}^{(1)[33]}$ & 1.6 & $B\to K\nu\bar\nu$ & 1.7 & $B\to K\nu\bar\nu$ & 6.1 & $B\to K\nu\bar\nu$ & 7.3 & $B\to K\nu\bar\nu$ \\
$\mathcal{C}_{\ell q}^{(3)[3333]}$ & -2.5 & $B\to K\nu\bar\nu$ & 1.5 & $B\to K\nu\bar\nu$ & 2.7 & $B\to K\nu\bar\nu$ & 7. & $B_s\to \tau\tau$ \\
$\mathcal{C}_{\ell q}^{(3)[ii33]}$ & -11.3 & $B\to K^{*}\nu\bar\nu$ & -5.5 & $B\to K\nu\bar\nu$ & 5.6 & $B\to K\nu\bar\nu$ & 6.3 & $B\to K\nu\bar\nu$ \\
$\mathcal{C}_{\ell q}^{(1)[33ii]}$ & -3.7 & $B\to K\nu\bar\nu$ & 1.7 & $B\to K\nu\bar\nu$ & 2.9 & $B\to K\nu\bar\nu$ & 6.2 & $B_s\to \tau\tau$ \\
$\mathcal{C}_{\ell q}^{(3)[33ii]}$ & -2.6 & $B\to K\nu\bar\nu$ & -1.4 & $B\to K\nu\bar\nu$ & 2.9 & $B\to K\nu\bar\nu$ & 6.2 & $B_s\to \tau\tau$ \\
$\mathcal{C}_{\ell q}^{(1)[3333]}$ & -7.2 & $B\to K^{*}\nu\bar\nu$ & -1.8 & $B\to K\nu\bar\nu$ & 2.9 & $B\to K\nu\bar\nu$ & 6.2 & $B_s\to \tau\tau$ \\
$\mathcal{C}_{\ell q}^{(1)[iijj]}$ & 5.5 & $B\to K^{*}\nu\bar\nu$ & 11. & $B\to K\nu\bar\nu$ & 5.7 & $B\to K\nu\bar\nu$ & 6.1 & $B\to K\nu\bar\nu$ \\
$\mathcal{C}_{\ell q}^{(1)[ii33]}$ & -10.9 & $B\to K\nu\bar\nu$ & -5.5 & $B\to K^{*}\nu\bar\nu$ & 5.6 & $B\to K\nu\bar\nu$ & 6.1 & $B\to K\nu\bar\nu$ \\
$\mathcal{C}_{\ell q}^{(3)[iijj]}$ & 5.6 & $B\to K\nu\bar\nu$ & 11.4 & $B\to K^{*}\nu\bar\nu$ & 5.6 & $B\to K\nu\bar\nu$ & 6. & $B\to K\nu\bar\nu$ \\
$\mathcal{C}_{qe}^{[ii33]}$ & -0.5 & $B_s\to \tau\tau$ & -0.3 & $B_s\to \tau\tau$ & 0.4 & $B_s\to \tau\tau$ & 5.4 & $B_s\to \tau\tau$ \\
$\mathcal{C}_{qe}^{[3333]}$ & 0.3 & $B_s\to \tau\tau$ & 0.5 & $B_s\to \tau\tau$ & 0.4 & $B_s\to \tau\tau$ & 5.3 & $B_s\to \tau\tau$ \\
$\mathcal{C}_{\ell \ell }^{[ii33]}$ & -0.9 & $\tau$ LFU & 0.9 & $\tau$ LFU & 0.9 & $\tau$ LFU & 4.1 & $\tau$ LFU \\
$\mathcal{C}_{\ell \ell }^{[iijj]}$ & -0.9 & $\tau$ LFU & 0.9 & $\tau$ LFU & 0.9 & $\tau$ LFU & 4.1 & $\tau$ LFU \\
$\mathcal{C}_{qu}^{(1)[3333]}$ & -3.1 & $B_s$ mixing & 1.7 & $B_s$ mixing & 2. & $B\to K\nu\bar\nu$ & 2.7 & $B\to K\nu\bar\nu$ \\
$\mathcal{C}_{qu}^{(1)[ii33]}$ & -1.7 & $B\to K\nu\bar\nu$ & 3.2 & $B_s$ mixing & 1.9 & $B\to K\nu\bar\nu$ & 2.6 & $B\to K\nu\bar\nu$ \\
$\mathcal{C}_{qu}^{(8)[3333]}$ & -0.8 & $B_s$ mixing & 0.8 & $B_s$ mixing & 1.2 & $B_s$ mixing & 1.7 & $B_s$ mixing \\
$\mathcal{C}_{qu}^{(8)[ii33]}$ & -0.6 & $B_s$ mixing & 0.8 & $B_s$ mixing & 1. & $B_s$ mixing & 1.5 & $B_s$ mixing \\
 \hline
\end{tabular}

%% file: main.bbl
\providecommand{\href}[2]{#2}\begingroup\raggedright\begin{thebibliography}{10}

\bibitem{Davighi:2023iks}
J.~Davighi and G.~Isidori, \emph{{Non-universal gauge interactions addressing
  the inescapable link between Higgs and flavour}},
  \href{https://doi.org/10.1007/JHEP07(2023)147}{\emph{JHEP} {\bfseries 07}
  (2023) 147} [\href{https://arxiv.org/abs/2303.01520}{{\ttfamily
  2303.01520}}].

\bibitem{Allwicher:2023shc}
L.~Allwicher, C.~Cornella, G.~Isidori and B.A.~Stefanek, \emph{{New physics in
  the third generation. A comprehensive SMEFT analysis and future prospects}},
  \href{https://doi.org/10.1007/JHEP03(2024)049}{\emph{JHEP} {\bfseries 03}
  (2024) 049} [\href{https://arxiv.org/abs/2311.00020}{{\ttfamily
  2311.00020}}].

\bibitem{Glioti:2024hye}
A.~Glioti, R.~Rattazzi, L.~Ricci and L.~Vecchi, \emph{{Exploring the Flavor
  Symmetry Landscape}},  \href{https://arxiv.org/abs/2402.09503}{{\ttfamily
  2402.09503}}.

\bibitem{FCC:2018evy}
{\scshape FCC} collaboration, \emph{{FCC-ee: The Lepton Collider}: {Future
  Circular Collider Conceptual Design Report Volume 2}},
  \href{https://doi.org/10.1140/epjst/e2019-900045-4}{\emph{Eur. Phys. J. ST}
  {\bfseries 228} (2019) 261}.

\bibitem{CEPCStudyGroup:2018ghi}
{\scshape CEPC Study Group} collaboration, \emph{{CEPC Conceptual Design
  Report: Volume 2 - Physics \& Detector}},
  \href{https://arxiv.org/abs/1811.10545}{{\ttfamily 1811.10545}}.

\bibitem{Monteil:2021ith}
S.~Monteil and G.~Wilkinson, \emph{{Heavy-quark opportunities and challenges at
  FCC-ee}}, \href{https://doi.org/10.1140/epjp/s13360-021-01814-0}{\emph{Eur.
  Phys. J. Plus} {\bfseries 136} (2021) 837}
  [\href{https://arxiv.org/abs/2106.01259}{{\ttfamily 2106.01259}}].

\bibitem{Amhis:2023mpj}
Y.~Amhis, M.~Kenzie, M.~Reboud and A.R.~Wiederhold, \emph{{Prospects for
  searches of $ b\to s\nu \overline{\nu} $ decays at FCC-ee}},
  \href{https://doi.org/10.1007/JHEP01(2024)144}{\emph{JHEP} {\bfseries 01}
  (2024) 144} [\href{https://arxiv.org/abs/2309.11353}{{\ttfamily
  2309.11353}}].

\bibitem{Zuo:2023dzn}
X.~Zuo, M.~Fedele, C.~Helsens, D.~Hill, S.~Iguro and M.~Klute, \emph{{Prospects
  for $B_c^+$ and $B^+\rightarrow \tau ^+ \nu _\tau $ at FCC-ee}},
  \href{https://doi.org/10.1140/epjc/s10052-024-12418-0}{\emph{Eur. Phys. J. C}
  {\bfseries 84} (2024) 87} [\href{https://arxiv.org/abs/2305.02998}{{\ttfamily
  2305.02998}}].

\bibitem{Ai:2024nmn}
X.~Ai et~al., \emph{{Flavor Physics at CEPC: a General Perspective}},
  \href{https://arxiv.org/abs/2412.19743}{{\ttfamily 2412.19743}}.

\bibitem{Barbieri:2011ci}
R.~Barbieri, G.~Isidori, J.~Jones-Perez, P.~Lodone and D.M.~Straub,
  \emph{{$U(2)$ and Minimal Flavour Violation in Supersymmetry}},
  \href{https://doi.org/10.1140/epjc/s10052-011-1725-z}{\emph{Eur. Phys. J. C}
  {\bfseries 71} (2011) 1725}
  [\href{https://arxiv.org/abs/1105.2296}{{\ttfamily 1105.2296}}].

\bibitem{Isidori:2012ts}
G.~Isidori and D.M.~Straub, \emph{{Minimal Flavour Violation and Beyond}},
  \href{https://doi.org/10.1140/epjc/s10052-012-2103-1}{\emph{Eur. Phys. J. C}
  {\bfseries 72} (2012) 2103}
  [\href{https://arxiv.org/abs/1202.0464}{{\ttfamily 1202.0464}}].

\bibitem{Faroughy:2020ina}
D.A.~Faroughy, G.~Isidori, F.~Wilsch and K.~Yamamoto, \emph{{Flavour symmetries
  in the SMEFT}}, \href{https://doi.org/10.1007/JHEP08(2020)166}{\emph{JHEP}
  {\bfseries 08} (2020) 166}
  [\href{https://arxiv.org/abs/2005.05366}{{\ttfamily 2005.05366}}].

\bibitem{Allwicher:2024ncl}
L.~Allwicher, M.~Bordone, G.~Isidori, G.~Piazza and A.~Stanzione,
  \emph{{Probing third-generation New Physics with $K\to \pi \nu\bar\nu$ and
  $B\to K^{(*)} \nu\bar\nu$}},
  \href{https://arxiv.org/abs/2410.21444}{{\ttfamily 2410.21444}}.

\bibitem{ParticleDataGroup:2024cfk}
{\scshape Particle Data Group} collaboration, \emph{{Review of particle
  physics}}, \href{https://doi.org/10.1103/PhysRevD.110.030001}{\emph{Phys.
  Rev. D} {\bfseries 110} (2024) 030001}.

\bibitem{Lusiani:2023tau}
A.~Lusiani, \emph{{Tau Physics Prospects at FCC-ee}},
  \href{https://doi.org/https://doi.org/10.17181/9bkm6-h8906}{\emph{CERN Note}
  }.

\bibitem{Belle-II:2018jsg}
{\scshape Belle-II} collaboration, \emph{{The Belle II Physics Book}},
  \href{https://doi.org/10.1093/ptep/ptz106}{\emph{PTEP} {\bfseries 2019}
  (2019) 123C01} [\href{https://arxiv.org/abs/1808.10567}{{\ttfamily
  1808.10567}}].

\bibitem{HeavyFlavorAveragingGroupHFLAV:2024ctg}
{\scshape Heavy Flavor Averaging Group (HFLAV)} collaboration, \emph{{Averages
  of $b$-hadron, $c$-hadron, and $\tau$-lepton properties as of 2023}},
  \href{https://arxiv.org/abs/2411.18639}{{\ttfamily 2411.18639}}.

\bibitem{Miralles:2024iii}
T.~Miralles, \emph{{Study of hadronic $B$ meson decays with the LHCb
  spectrometer (at LHC) and an investigation of detector requirements for the
  Future Circular Collider (FCC- $ee$)}},  other thesis, 2024, 2024.

\bibitem{Charles:2020dfl}
J.~Charles, S.~Descotes-Genon, Z.~Ligeti, S.~Monteil, M.~Papucci, K.~Trabelsi
  et~al., \emph{{New physics in $B$ meson mixing: future sensitivity and
  limitations}}, \href{https://doi.org/10.1103/PhysRevD.102.056023}{\emph{Phys.
  Rev. D} {\bfseries 102} (2020) 056023}
  [\href{https://arxiv.org/abs/2006.04824}{{\ttfamily 2006.04824}}].

\bibitem{LHCb:2018roe}
{\scshape LHCb} collaboration, \emph{{Physics case for an LHCb Upgrade II -
  Opportunities in flavour physics, and beyond, in the HL-LHC era}},
  \href{https://arxiv.org/abs/1808.08865}{{\ttfamily 1808.08865}}.

\bibitem{deBlas:2022ofj}
J.~de~Blas, Y.~Du, C.~Grojean, J.~Gu, V.~Miralles, M.E.~Peskin et~al.,
  \emph{{Global SMEFT Fits at Future Colliders}},  in \emph{{Snowmass 2021}},
  6, 2022 [\href{https://arxiv.org/abs/2206.08326}{{\ttfamily 2206.08326}}].

\bibitem{Blondel:2021ema}
A.~Blondel and P.~Janot, \emph{{FCC-ee overview: new opportunities create new
  challenges}},
  \href{https://doi.org/10.1140/epjp/s13360-021-02154-9}{\emph{Eur. Phys. J.
  Plus} {\bfseries 137} (2022) 92}
  [\href{https://arxiv.org/abs/2106.13885}{{\ttfamily 2106.13885}}].

\bibitem{Bernardi:2022hny}
G.~Bernardi et~al., \emph{{The Future Circular Collider: a Summary for the US
  2021 Snowmass Process}},  \href{https://arxiv.org/abs/2203.06520}{{\ttfamily
  2203.06520}}.

\bibitem{Buchmuller:1985jz}
W.~Buchmuller and D.~Wyler, \emph{{Effective Lagrangian Analysis of New
  Interactions and Flavor Conservation}},
  \href{https://doi.org/10.1016/0550-3213(86)90262-2}{\emph{Nucl. Phys. B}
  {\bfseries 268} (1986) 621}.

\bibitem{Grzadkowski:2010es}
B.~Grzadkowski, M.~Iskrzynski, M.~Misiak and J.~Rosiek, \emph{{Dimension-Six
  Terms in the Standard Model Lagrangian}},
  \href{https://doi.org/10.1007/JHEP10(2010)085}{\emph{JHEP} {\bfseries 10}
  (2010) 085} [\href{https://arxiv.org/abs/1008.4884}{{\ttfamily 1008.4884}}].

\bibitem{Brivio:2017vri}
I.~Brivio and M.~Trott, \emph{{The Standard Model as an Effective Field
  Theory}}, \href{https://doi.org/10.1016/j.physrep.2018.11.002}{\emph{Phys.
  Rept.} {\bfseries 793} (2019) 1}
  [\href{https://arxiv.org/abs/1706.08945}{{\ttfamily 1706.08945}}].

\bibitem{Isidori:2023pyp}
G.~Isidori, F.~Wilsch and D.~Wyler, \emph{{The standard model effective field
  theory at work}},
  \href{https://doi.org/10.1103/RevModPhys.96.015006}{\emph{Rev. Mod. Phys.}
  {\bfseries 96} (2024) 015006}
  [\href{https://arxiv.org/abs/2303.16922}{{\ttfamily 2303.16922}}].

\bibitem{Dvali:2000ha}
G.R.~Dvali and M.A.~Shifman, \emph{{Families as neighbors in extra dimension}},
  \href{https://doi.org/10.1016/S0370-2693(00)00083-6}{\emph{Phys. Lett. B}
  {\bfseries 475} (2000) 295}
  [\href{https://arxiv.org/abs/hep-ph/0001072}{{\ttfamily hep-ph/0001072}}].

\bibitem{Panico:2016ull}
G.~Panico and A.~Pomarol, \emph{{Flavor hierarchies from dynamical scales}},
  \href{https://doi.org/10.1007/JHEP07(2016)097}{\emph{JHEP} {\bfseries 07}
  (2016) 097} [\href{https://arxiv.org/abs/1603.06609}{{\ttfamily
  1603.06609}}].

\bibitem{Barbieri:2021wrc}
R.~Barbieri, \emph{{A View of Flavour Physics in 2021}},
  \href{https://doi.org/10.5506/APhysPolB.52.789}{\emph{Acta Phys. Polon. B}
  {\bfseries 52} (2021) 789}
  [\href{https://arxiv.org/abs/2103.15635}{{\ttfamily 2103.15635}}].

\bibitem{Allwicher:2020esa}
L.~Allwicher, G.~Isidori and A.E.~Thomsen, \emph{{Stability of the Higgs Sector
  in a Flavor-Inspired Multi-Scale Model}},
  \href{https://doi.org/10.1007/JHEP01(2021)191}{\emph{JHEP} {\bfseries 01}
  (2021) 191} [\href{https://arxiv.org/abs/2011.01946}{{\ttfamily
  2011.01946}}].

\bibitem{Bordone:2017bld}
M.~Bordone, C.~Cornella, J.~Fuentes-Martin and G.~Isidori, \emph{{A three-site
  gauge model for flavor hierarchies and flavor anomalies}},
  \href{https://doi.org/10.1016/j.physletb.2018.02.011}{\emph{Phys. Lett. B}
  {\bfseries 779} (2018) 317}
  [\href{https://arxiv.org/abs/1712.01368}{{\ttfamily 1712.01368}}].

\bibitem{Greljo:2018tuh}
A.~Greljo and B.A.~Stefanek, \emph{{Third family quark\textendash{}lepton
  unification at the TeV scale}},
  \href{https://doi.org/10.1016/j.physletb.2018.05.033}{\emph{Phys. Lett. B}
  {\bfseries 782} (2018) 131}
  [\href{https://arxiv.org/abs/1802.04274}{{\ttfamily 1802.04274}}].

\bibitem{Fuentes-Martin:2020pww}
J.~Fuentes-Martin, G.~Isidori, J.~Pag\`es and B.A.~Stefanek, \emph{{Flavor
  non-universal Pati-Salam unification and neutrino masses}},
  \href{https://doi.org/10.1016/j.physletb.2021.136484}{\emph{Phys. Lett. B}
  {\bfseries 820} (2021) 136484}
  [\href{https://arxiv.org/abs/2012.10492}{{\ttfamily 2012.10492}}].

\bibitem{Fuentes-Martin:2022xnb}
J.~Fuentes-Martin, G.~Isidori, J.M.~Lizana, N.~Selimovic and B.A.~Stefanek,
  \emph{{Flavor hierarchies, flavor anomalies, and Higgs mass from a warped
  extra dimension}},
  \href{https://doi.org/10.1016/j.physletb.2022.137382}{\emph{Phys. Lett. B}
  {\bfseries 834} (2022) 137382}
  [\href{https://arxiv.org/abs/2203.01952}{{\ttfamily 2203.01952}}].

\bibitem{Davighi:2022bqf}
J.~Davighi, G.~Isidori and M.~Pesut, \emph{{Electroweak-flavour and
  quark-lepton unification: a family non-universal path}},
  \href{https://doi.org/10.1007/JHEP04(2023)030}{\emph{JHEP} {\bfseries 04}
  (2023) 030} [\href{https://arxiv.org/abs/2212.06163}{{\ttfamily
  2212.06163}}].

\bibitem{Davighi:2023evx}
J.~Davighi and B.A.~Stefanek, \emph{{Deconstructed hypercharge: a natural model
  of flavour}}, \href{https://doi.org/10.1007/JHEP11(2023)100}{\emph{JHEP}
  {\bfseries 11} (2023) 100}
  [\href{https://arxiv.org/abs/2305.16280}{{\ttfamily 2305.16280}}].

\bibitem{Barbieri:2023qpf}
R.~Barbieri and G.~Isidori, \emph{{Minimal flavour deconstruction}},
  \href{https://doi.org/10.1007/JHEP05(2024)033}{\emph{JHEP} {\bfseries 05}
  (2024) 033} [\href{https://arxiv.org/abs/2312.14004}{{\ttfamily
  2312.14004}}].

\bibitem{FernandezNavarro:2023rhv}
M.~Fern\'andez~Navarro and S.F.~King, \emph{{Tri-hypercharge: a separate gauged
  weak hypercharge for each fermion family as the origin of flavour}},
  \href{https://doi.org/10.1007/JHEP08(2023)020}{\emph{JHEP} {\bfseries 08}
  (2023) 020} [\href{https://arxiv.org/abs/2305.07690}{{\ttfamily
  2305.07690}}].

\bibitem{FernandezNavarro:2024hnv}
M.~Fern\'andez~Navarro, S.F.~King and A.~Vicente, \emph{{Minimal complete
  tri-hypercharge theories of flavour}},
  \href{https://doi.org/10.1007/JHEP07(2024)147}{\emph{JHEP} {\bfseries 07}
  (2024) 147} [\href{https://arxiv.org/abs/2404.12442}{{\ttfamily
  2404.12442}}].

\bibitem{Stefanek:2024kds}
B.A.~Stefanek, \emph{{Non-universal probes of composite Higgs models: new
  bounds and prospects for FCC-ee}},
  \href{https://doi.org/10.1007/JHEP09(2024)103}{\emph{JHEP} {\bfseries 09}
  (2024) 103} [\href{https://arxiv.org/abs/2407.09593}{{\ttfamily
  2407.09593}}].

\bibitem{Covone:2024elw}
S.~Covone, J.~Davighi, G.~Isidori and M.~Pesut, \emph{{Flavour deconstructing
  the composite Higgs}},
  \href{https://doi.org/10.1007/JHEP01(2025)041}{\emph{JHEP} {\bfseries 01}
  (2025) 041} [\href{https://arxiv.org/abs/2407.10950}{{\ttfamily
  2407.10950}}].

\bibitem{Iguro:2024hyk}
S.~Iguro, T.~Kitahara and R.~Watanabe, \emph{{Global fit to
  b\textrightarrow{}c\ensuremath{\tau}\ensuremath{\nu} anomalies as of Spring
  2024}}, \href{https://doi.org/10.1103/PhysRevD.110.075005}{\emph{Phys. Rev.
  D} {\bfseries 110} (2024) 075005}
  [\href{https://arxiv.org/abs/2405.06062}{{\ttfamily 2405.06062}}].

\bibitem{Belle-II:2023esi}
{\scshape Belle-II} collaboration, \emph{{Evidence for
  B+\textrightarrow{}K+\ensuremath{\nu}\ensuremath{\nu}\textasciimacron{}
  decays}}, \href{https://doi.org/10.1103/PhysRevD.109.112006}{\emph{Phys. Rev.
  D} {\bfseries 109} (2024) 112006}
  [\href{https://arxiv.org/abs/2311.14647}{{\ttfamily 2311.14647}}].

\bibitem{NA62:2024pjp}
{\scshape NA62} collaboration, \emph{{Observation of the
  $K^{+}\rightarrow\pi^{+}\nu\bar{\nu}$ decay and measurement of its branching
  ratio}},  \href{https://arxiv.org/abs/2412.12015}{{\ttfamily 2412.12015}}.

\bibitem{Alguero:2022wkd}
M.~Alguer\'o, J.~Matias, B.~Capdevila and A.~Crivellin, \emph{{Disentangling
  lepton flavor universal and lepton flavor universality violating effects in
  b\textrightarrow{}s\ensuremath{\ell}+\ensuremath{\ell}- transitions}},
  \href{https://doi.org/10.1103/PhysRevD.105.113007}{\emph{Phys. Rev. D}
  {\bfseries 105} (2022) 113007}
  [\href{https://arxiv.org/abs/2205.15212}{{\ttfamily 2205.15212}}].

\bibitem{Bordone:2024hui}
M.~Bordone, G.~isidori, S.~M\"achler and A.~Tinari, \emph{{Short- vs.
  long-distance physics in $B\rightarrow K^{(*)} \ell ^+\ell ^-$: a data-driven
  analysis}}, \href{https://doi.org/10.1140/epjc/s10052-024-12869-5}{\emph{Eur.
  Phys. J. C} {\bfseries 84} (2024) 547}
  [\href{https://arxiv.org/abs/2401.18007}{{\ttfamily 2401.18007}}].

\bibitem{Isidori:2024lng}
G.~Isidori, Z.~Polonsky and A.~Tinari, \emph{{An explicit estimate of charm
  rescattering in $B^0 \to K^0 \bar{\ell} \ell$}},
  \href{https://arxiv.org/abs/2405.17551}{{\ttfamily 2405.17551}}.

\bibitem{Marzocca:2024hua}
D.~Marzocca, M.~Nardecchia, A.~Stanzione and C.~Toni, \emph{{Implications of $B
  \rightarrow K \nu {\bar{\nu }}$ under rank-one flavor violation hypothesis}},
  \href{https://doi.org/10.1140/epjc/s10052-024-13534-7}{\emph{Eur. Phys. J. C}
  {\bfseries 84} (2024) 1217}
  [\href{https://arxiv.org/abs/2404.06533}{{\ttfamily 2404.06533}}].

\bibitem{Alonso:2015sja}
R.~Alonso, B.~Grinstein and J.~Martin~Camalich, \emph{{Lepton universality
  violation and lepton flavor conservation in B-meson decays}},
  \href{https://doi.org/10.1007/JHEP10(2015)184}{\emph{JHEP} {\bfseries 10}
  (2015) 184} [\href{https://arxiv.org/abs/1505.05164}{{\ttfamily
  1505.05164}}].

\bibitem{Calibbi:2015kma}
L.~Calibbi, A.~Crivellin and T.~Ota, \emph{{Effective Field Theory Approach to
  $b\to s\ell\ell^{(')}$, $B\to K^{(*)}\nu\overline{\nu}$ and $B\to
  D^{(*)}\tau\nu$ with Third Generation Couplings}},
  \href{https://doi.org/10.1103/PhysRevLett.115.181801}{\emph{Phys. Rev. Lett.}
  {\bfseries 115} (2015) 181801}
  [\href{https://arxiv.org/abs/1506.02661}{{\ttfamily 1506.02661}}].

\bibitem{Barbieri:2015yvd}
R.~Barbieri, G.~Isidori, A.~Pattori and F.~Senia, \emph{{Anomalies in B-decays
  and U (2) flavour symmetry}},
  \href{https://doi.org/10.1140/epjc/s10052-016-3905-3}{\emph{Eur. Phys. J. C}
  {\bfseries 76} (2016) 67} [\href{https://arxiv.org/abs/1512.01560}{{\ttfamily
  1512.01560}}].

\bibitem{Bhattacharya:2016mcc}
B.~Bhattacharya, A.~Datta, J.-P.~Gu\'evin, D.~London and R.~Watanabe,
  \emph{{Simultaneous Explanation of the RK and RD(*) Puzzles: a Model
  Analysis}}, \href{https://doi.org/10.1007/JHEP01(2017)015}{\emph{JHEP}
  {\bfseries 01} (2017) 015}
  [\href{https://arxiv.org/abs/1609.09078}{{\ttfamily 1609.09078}}].

\bibitem{Buttazzo:2017ixm}
D.~Buttazzo, A.~Greljo, G.~Isidori and D.~Marzocca, \emph{{B-physics anomalies:
  a guide to combined explanations}},
  \href{https://doi.org/10.1007/JHEP11(2017)044}{\emph{JHEP} {\bfseries 11}
  (2017) 044} [\href{https://arxiv.org/abs/1706.07808}{{\ttfamily
  1706.07808}}].

\bibitem{DiLuzio:2017vat}
L.~Di~Luzio, A.~Greljo and M.~Nardecchia, \emph{{Gauge leptoquark as the origin
  of B-physics anomalies}},
  \href{https://doi.org/10.1103/PhysRevD.96.115011}{\emph{Phys. Rev. D}
  {\bfseries 96} (2017) 115011}
  [\href{https://arxiv.org/abs/1708.08450}{{\ttfamily 1708.08450}}].

\bibitem{Baker:2019sli}
M.J.~Baker, J.~Fuentes-Mart\'\i{}n, G.~Isidori and M.~K\"onig, \emph{{High-
  $p_T$ signatures in vector\textendash{}leptoquark models}},
  \href{https://doi.org/10.1140/epjc/s10052-019-6853-x}{\emph{Eur. Phys. J. C}
  {\bfseries 79} (2019) 334}
  [\href{https://arxiv.org/abs/1901.10480}{{\ttfamily 1901.10480}}].

\bibitem{Pati:1974yy}
J.C.~Pati and A.~Salam, \emph{{Lepton Number as the Fourth Color}},
  \href{https://doi.org/10.1103/PhysRevD.10.275}{\emph{Phys. Rev. D} {\bfseries
  10} (1974) 275}.

\bibitem{DiLuzio:2018zxy}
L.~Di~Luzio, J.~Fuentes-Martin, A.~Greljo, M.~Nardecchia and S.~Renner,
  \emph{{Maximal Flavour Violation: a Cabibbo mechanism for leptoquarks}},
  \href{https://doi.org/10.1007/JHEP11(2018)081}{\emph{JHEP} {\bfseries 11}
  (2018) 081} [\href{https://arxiv.org/abs/1808.00942}{{\ttfamily
  1808.00942}}].

\bibitem{Aebischer:2022oqe}
J.~Aebischer, G.~Isidori, M.~Pesut, B.A.~Stefanek and F.~Wilsch,
  \emph{{Confronting the vector leptoquark hypothesis with new low- and
  high-energy data}},
  \href{https://doi.org/10.1140/epjc/s10052-023-11304-5}{\emph{Eur. Phys. J. C}
  {\bfseries 83} (2023) 153}
  [\href{https://arxiv.org/abs/2210.13422}{{\ttfamily 2210.13422}}].

\bibitem{Fuentes-Martin:2019ign}
J.~Fuentes-Mart\'\i{}n, G.~Isidori, M.~K\"onig and N.~Selimovi\'c,
  \emph{{Vector Leptoquarks Beyond Tree Level}},
  \href{https://doi.org/10.1103/PhysRevD.101.035024}{\emph{Phys. Rev. D}
  {\bfseries 101} (2020) 035024}
  [\href{https://arxiv.org/abs/1910.13474}{{\ttfamily 1910.13474}}].

\bibitem{Fuentes-Martin:2020luw}
J.~Fuentes-Mart\'\i{}n, G.~Isidori, M.~K\"onig and N.~Selimovi\'c,
  \emph{{Vector leptoquarks beyond tree level. II. $\mathcal{O}(\alpha_s)$
  corrections and radial modes}},
  \href{https://doi.org/10.1103/PhysRevD.102.035021}{\emph{Phys. Rev. D}
  {\bfseries 102} (2020) 035021}
  [\href{https://arxiv.org/abs/2006.16250}{{\ttfamily 2006.16250}}].

\bibitem{Fuentes-Martin:2020hvc}
J.~Fuentes-Mart\'\i{}n, G.~Isidori, M.~K\"onig and N.~Selimovi\'c,
  \emph{{Vector Leptoquarks Beyond Tree Level III: Vector-like Fermions and
  Flavor-Changing Transitions}},
  \href{https://doi.org/10.1103/PhysRevD.102.115015}{\emph{Phys. Rev. D}
  {\bfseries 102} (2020) 115015}
  [\href{https://arxiv.org/abs/2009.11296}{{\ttfamily 2009.11296}}].

\bibitem{Bauer:2015knc}
M.~Bauer and M.~Neubert, \emph{{Minimal Leptoquark Explanation for the
  $R_{D^{(*)}}$ , $R_K$ , and $(g-2)_\mu$ Anomalies}},
  \href{https://doi.org/10.1103/PhysRevLett.116.141802}{\emph{Phys. Rev. Lett.}
  {\bfseries 116} (2016) 141802}
  [\href{https://arxiv.org/abs/1511.01900}{{\ttfamily 1511.01900}}].

\bibitem{Crivellin:2017zlb}
A.~Crivellin, D.~M\"uller and T.~Ota, \emph{{Simultaneous explanation of
  $R(D^{(*)})$ and $b\to s \mu^+\mu^-$: the last scalar leptoquarks standing}},
  \href{https://doi.org/10.1007/JHEP09(2017)040}{\emph{JHEP} {\bfseries 09}
  (2017) 040} [\href{https://arxiv.org/abs/1703.09226}{{\ttfamily
  1703.09226}}].

\bibitem{Gherardi:2020det}
V.~Gherardi, D.~Marzocca and E.~Venturini, \emph{{Matching scalar leptoquarks
  to the SMEFT at one loop}},
  \href{https://doi.org/10.1007/JHEP07(2020)225}{\emph{JHEP} {\bfseries 07}
  (2020) 225} [\href{https://arxiv.org/abs/2003.12525}{{\ttfamily
  2003.12525}}].

\bibitem{Ishiwata:2015cga}
K.~Ishiwata, Z.~Ligeti and M.B.~Wise, \emph{{New Vector-Like Fermions and
  Flavor Physics}}, \href{https://doi.org/10.1007/JHEP10(2015)027}{\emph{JHEP}
  {\bfseries 10} (2015) 027}
  [\href{https://arxiv.org/abs/1506.03484}{{\ttfamily 1506.03484}}].

\bibitem{Bobeth:2016llm}
C.~Bobeth, A.J.~Buras, A.~Celis and M.~Jung, \emph{{Patterns of Flavour
  Violation in Models with Vector-Like Quarks}},
  \href{https://doi.org/10.1007/JHEP04(2017)079}{\emph{JHEP} {\bfseries 04}
  (2017) 079} [\href{https://arxiv.org/abs/1609.04783}{{\ttfamily
  1609.04783}}].

\bibitem{Mann:2017wzh}
K.~Ishiwata, Z.~Ligeti and M.B.~Wise, \emph{{New Vector-Like Fermions and
  Flavor Physics}}, \href{https://doi.org/10.1007/JHEP10(2015)027}{\emph{JHEP}
  {\bfseries 10} (2015) 027}
  [\href{https://arxiv.org/abs/1506.03484}{{\ttfamily 1506.03484}}].

\bibitem{Alves:2023ufm}
J.a.M.~Alves, G.C.~Branco, A.L.~Cherchiglia, C.C.~Nishi, J.T.~Penedo,
  P.M.F.~Pereira et~al., \emph{{Vector-like singlet quarks: A roadmap}},
  \href{https://doi.org/10.1016/j.physrep.2023.12.004}{\emph{Phys. Rept.}
  {\bfseries 1057} (2024) 1}
  [\href{https://arxiv.org/abs/2304.10561}{{\ttfamily 2304.10561}}].

\bibitem{Allwicher:2022mcg}
L.~Allwicher, D.A.~Faroughy, F.~Jaffredo, O.~Sumensari and F.~Wilsch,
  \emph{{HighPT: A tool for~ high-$p_T$ Drell-Yan tails beyond the standard
  model}}, \href{https://doi.org/10.1016/j.cpc.2023.108749}{\emph{Comput. Phys.
  Commun.} {\bfseries 289} (2023) 108749}
  [\href{https://arxiv.org/abs/2207.10756}{{\ttfamily 2207.10756}}].

\bibitem{ALEPH:2013dgf}
{\scshape ALEPH, DELPHI, L3, OPAL, LEP Electroweak} collaboration,
  \emph{{Electroweak Measurements in Electron-Positron Collisions at
  W-Boson-Pair Energies at LEP}},
  \href{https://doi.org/10.1016/j.physrep.2013.07.004}{\emph{Phys. Rept.}
  {\bfseries 532} (2013) 119}
  [\href{https://arxiv.org/abs/1302.3415}{{\ttfamily 1302.3415}}].

\bibitem{Hartland:2019bjb}
N.P.~Hartland, F.~Maltoni, E.R.~Nocera, J.~Rojo, E.~Slade, E.~Vryonidou et~al.,
  \emph{{A Monte Carlo global analysis of the Standard Model Effective Field
  Theory: the top quark sector}},
  \href{https://doi.org/10.1007/JHEP04(2019)100}{\emph{JHEP} {\bfseries 04}
  (2019) 100} [\href{https://arxiv.org/abs/1901.05965}{{\ttfamily
  1901.05965}}].

\bibitem{Ethier:2021bye}
{\scshape SMEFiT} collaboration, \emph{{Combined SMEFT interpretation of Higgs,
  diboson, and top quark data from the LHC}},
  \href{https://doi.org/10.1007/JHEP11(2021)089}{\emph{JHEP} {\bfseries 11}
  (2021) 089} [\href{https://arxiv.org/abs/2105.00006}{{\ttfamily
  2105.00006}}].

\bibitem{Breso-Pla:2021qoe}
V.~Bres\'o-Pla, A.~Falkowski and M.~Gonz\'alez-Alonso, \emph{{A$_{FB}$ in the
  SMEFT: precision Z physics at the LHC}},
  \href{https://doi.org/10.1007/JHEP08(2021)021}{\emph{JHEP} {\bfseries 08}
  (2021) 021} [\href{https://arxiv.org/abs/2103.12074}{{\ttfamily
  2103.12074}}].

\bibitem{Fuentes-Martin:2020zaz}
J.~Fuentes-Martin, P.~Ruiz-Femenia, A.~Vicente and J.~Virto, \emph{{DsixTools
  2.0: The Effective Field Theory Toolkit}},
  \href{https://doi.org/10.1140/epjc/s10052-020-08778-y}{\emph{Eur. Phys. J. C}
  {\bfseries 81} (2021) 167}
  [\href{https://arxiv.org/abs/2010.16341}{{\ttfamily 2010.16341}}].

\bibitem{Greljo:2024ytg}
A.~Greljo, H.~Tiblom and A.~Valenti, \emph{{New Physics Through Flavor Tagging
  at FCC-ee}},  \href{https://arxiv.org/abs/2411.02485}{{\ttfamily
  2411.02485}}.

\bibitem{Allwicher:2021ndi}
L.~Allwicher, G.~Isidori and N.~Selimovic, \emph{{LFU violations in leptonic
  \ensuremath{\tau} decays and B-physics anomalies}},
  \href{https://doi.org/10.1016/j.physletb.2022.136903}{\emph{Phys. Lett. B}
  {\bfseries 826} (2022) 136903}
  [\href{https://arxiv.org/abs/2109.03833}{{\ttfamily 2109.03833}}].

\bibitem{Cornella:2021sby}
C.~Cornella, D.A.~Faroughy, J.~Fuentes-Martin, G.~Isidori and M.~Neubert,
  \emph{{Reading the footprints of the B-meson flavor anomalies}},
  \href{https://doi.org/10.1007/JHEP08(2021)050}{\emph{JHEP} {\bfseries 08}
  (2021) 050} [\href{https://arxiv.org/abs/2103.16558}{{\ttfamily
  2103.16558}}].

\bibitem{Allwicher:2022gkm}
L.~Allwicher, D.A.~Faroughy, F.~Jaffredo, O.~Sumensari and F.~Wilsch,
  \emph{{Drell-Yan tails beyond the Standard Model}},
  \href{https://doi.org/10.1007/JHEP03(2023)064}{\emph{JHEP} {\bfseries 03}
  (2023) 064} [\href{https://arxiv.org/abs/2207.10714}{{\ttfamily
  2207.10714}}].

\bibitem{Allwicher:2023aql}
L.~Allwicher, G.~Isidori, J.M.~Lizana, N.~Selimovic and B.A.~Stefanek,
  \emph{{Third-family quark-lepton Unification and electroweak precision
  tests}}, \href{https://doi.org/10.1007/JHEP05(2023)179}{\emph{JHEP}
  {\bfseries 05} (2023) 179}
  [\href{https://arxiv.org/abs/2302.11584}{{\ttfamily 2302.11584}}].

\end{thebibliography}\endgroup
